\documentclass[a4paper,fleqn,usenatbib]{mnras}

\usepackage{newtxtext,newtxmath}
\usepackage{physics}
\usepackage{wasysym}
\usepackage[T1]{fontenc}
\usepackage{ae,aecompl}

\usepackage{graphicx}	
\usepackage{amsmath}	
\usepackage{amssymb}	
\usepackage{mwe}

\title[Photoionization in accretion torus]{Massive star formation via torus accretion: the effect of photoionization feedback}

\author[N. S. Sartorio et al.]{
N. S. Sartorio,$^{1}$\thanks{E-mail: nina.sartorio@gmail.com (NSS)}
B. Vandenbroucke,$^{2}$
D. Falceta-Goncalves$^{1,3}$
, K. Wood$^{2}$ and E. Keto$^{4}$
\\
$^{1}$Instituto Nacional de Pesquisas Espaciais - INPE, Divis\~ao de Astrof\'isica, Av. dos Astronautas, 1.758 -Jardim da Granja, S\~ao Jos\'e dos Campos - SP, Brazil\\
$^{2}$ SUPA, School of Physics \& Astronomy, University of St Andrews, North Haugh, St Andrews, KY16 9SS, United Kingdom\\
$^{3}$Escola de Artes, Ci\^encias e Humanidades, Universidade de S\~ao Paulo, Rua Arlindo Bettio 1000, S\~ao Paulo, SP, 03828-000, Brazil\\
$^{4}$ Harvard-Smithsonian Center for Astrophysics, 160 Garden St, Cambridge, MA 02420, USA
}

\date{Accepted XXX. Received YYY; in original form ZZZ}

\pubyear{2018}

\begin{document}
\label{firstpage}
\pagerange{\pageref{firstpage}--\pageref{lastpage}}
\maketitle

\begin{abstract}
The formation of massive stars is a long standing problem. Although a number of theories of massive star formation exist, ideas appear to converge to a disk-mediated accretion scenario. Here we present radiative hydrodynamic simulations of a star accreting mass via a disk embedded in a torus. We use a Monte Carlo based radiation hydrodynamics code to investigate the impact that ionizing radiation has on the torus. Ionized regions in the torus midplane are found to be either gravitationally trapped or in pressure driven expansion depending on whether or not the size of the ionized region exceeds a critical radius. Trapped H{\sc ii}  regions in the torus plane allow accretion to progress, while expanding H{\sc ii} regions disrupt the accretion torus preventing the central star from aggregating more mass, thereby setting the star's final mass. We obtain constraints for the luminosities and torus densities that lead to both scenarios. 
\end{abstract}




\section{Introduction}

Massive stars (from 10 up to above 100 solar masses) represent a minority of the stars in the stellar initial mass function \citep{Massey}. They have, however, a profound impact on their environment as their feedback mechanisms (winds, outflows, radiation pressure, photoionization, supernovae)  inject vast amounts of energy and momentum leading to dynamical and chemical changes in the surrounding gas \citep[e.g.][]{Kennicutt}.

Despite their importance, the formation of massive stars remains in many ways unclear. One of the reasons modelling massive star formation is challenging is that, in comparison with low mass stars, there is very little observational information \citep{beltran}. This is due to the fact that massive stars form quickly and hence the sites where ongoing massive star formation can be observed are rare. As a consequence of their rapid formation, massive proto-stars do not have time to clear out their surroundings and are, therefore, deeply concealed in the innermost region of an infalling envelope of gas and dust \citep{cesaroni}. This  makes mm and sub-mm wavelength observations the only way to probe deep into the core region \citep{beltran}. Observations are further complicated by the fact that the closest massive proto-stars are at least a kiloparsec away \citep{zinnecker, 2014.Tan.Review}. These large distances imply that sub-arcsecond resolution is required in the measurements if we are to resolve, for instance, disk structures.

Due to all these challenges at the observational front it is not clear how massive stars accrete. Early works considered spherically symmetric accretion onto the massive proto-stars \citep{2003.keto.spherical}, whereas more recent papers seem to have focused on to a disk accretion mode  \citep[e.g.][]{Haemmerle, jankovic, 2017harries, 2018.kuiper}. 

One of the main problems with spherical accretion is often the inability of the star to keep accreting past the 40 solar mass mark. This mass ceiling arises because radiative feedback eventually leads accretion to stop completely; thereby preventing the star acquiring more mass. One such feedback mechanism is radiation pressure on dust grains. A (proto)star could emit enough radiation such that the pressure on dust would be stronger than its gravitational pull, reversing the flow direction and preventing further accretion  \citep{kahn}. However, the maximum attainable mass of the star before this reversal happens depends highly on the dust grain properties that are being taken into account \citep{wolfire.cassinelli}.

Another mechanism that may shut down accretion is ionizing radiation. The Kelvin-Helmholtz timescale for a massive star is much shorter than the time it would take the star to reach its final mass at a realistic accretion rate \citep[e.g.][]{2009Krumholz}. As a result, in order to form, massive stars must continue accreting even after the onset of hydrogen fusion in their core, that is, when they join the zero age main sequence \citep[e.g.][]{zinnecker}. After the onset of fusion the massive star begins emitting considerable amounts of ionizing radiation and, therefore, we can expect the formation of an ionized (H{\sc ii}) region around it.

Classically, H{\sc ii} regions have been thought as pressure-driven bubbles expanding through space \citep{1978.spitzer}. Consider the simple scenario of a point source isotropically emitting at a certain ionizing luminosity, characterized by an emission rate of ionizing photons. The resulting ionization is counteracted by the rate of recombination of atoms to their neutral state, characterized by a recombination rate. The H{\sc ii} region will expand rapidly up to the point at which these two rates exactly counterbalance each other. In a spherically symmetric scenario, this would be referred to as a  Str\"{o}mgren sphere \citep{1939stromgren}. Although the H{\sc ii} region has a number density that is only twice that of the neutral gas, its temperature will be around two orders of magnitude higher. Therefore, the pressure inside the ionized bubble is much larger than in its surroundings and the ionized region will begin expanding into the neutral gas via a shock front. As it expands, its density decreases and thus the recombination rate falls, meaning that the same luminosity is able to ionize out to a larger radius. The expansion will continue until the bubble is able to re-establish a pressure equilibrium with the neutral region.

If H{\sc ii} regions behaved like that around forming massive (proto)stars, this would imply that accretion is no longer possible once the star starts emitting ionizing radiation. However, the works of \citet{Mestel}  and \citet{Keto_trapped1} showed that this simple picture changes dramatically if the gravity of the central point mass is taken into account. In this case, material crossing the ionization front is subject to two forces: the gravitational force, pulling the material inward, and the force due to the pressure gradient at the ionization front, pushing outwards. As long as the gravitational force is greater than the pressure gradient, the H{\sc ii} region will not be able to expand. This is referred to as a`trapped' H{\sc ii} region and allows material to keep flowing through the ionization front towards the central star. In other words, trapped H{\sc ii} regions make it possible for accretion to continue even after the star starts emitting significant amounts of ionizing photons.

The condition for a trapped H{\sc ii}  regions has been shown to be satisfied if its size does not exceed a critical radius roughly equal to the sonic radius (the point at which the escape velocity is the same as the sound speed).

However, as a star evolves and its ionizing luminosity increases, the H{\sc ii} region will increase in size. As a result, the trapped ionized region can eventually grow past this critical radius and evolve into an expanding H{\sc ii} region.
The existence of a trapped phase implies that a star could, in theory, extend its accretion period for as long as the trapped H{\sc ii} region phase lasts.

A perhaps more natural solution to the issues brought up by radiation pressure and photoionization is to assume that massive stars, as their low mass counterparts, accrete mainly via a disk. This is to be expected as accreting material is likely to have a tangential velocity component, and a disk is a natural mechanism that allows for the redistribution of angular momentum. Indeed, despite differing in the way large scale accretion occurs\footnote{In the core collapse model the maximum mass of the star is pre-determined by the size of the parent clump whereas in competitive accretion a number of stars form simultaneously in a common gravitational potential and stars located at the bottom of the potential can have much higher accretion rates.}, both models, Core Accretion \citep{turbulent.core} and Competitive Accretion  \citep{Bonnell.and.Bate.2006, 2001.Bonnell}, seem to agree that accretion onto the protostar probably takes place via a disk \citep{2014.Tan.Review}. Furthermore, an increasing body of observations of tori and disks around forming massive stars has been building up, even though the current evidence for disks limits itself to stars below 30 solar masses \citep[for a review, see][]{beltran}.

There are several numerical simulations of massive star formation, but due to the complex nature of the problem they vary significantly in what physics is included, usually containing one or more of the following factors: radiation pressure, photoionization, magnetic fields, outflows and stellar evolution models \citep[see for instance ][]{2017harries,2018.kuiper}. These simulations tend to focus on either the effect of one or two physical processes in 3D or they include the impact of many processes but compromise in resolution and/or make the simulations 2D by assuming additional symmetries for the system.

In this paper we choose to focus on the impact of ionization feedback effects on a massive star accreting via a disk embedded in a torus.  \citet{2007Keto}, predicted the existence of three stages in the evolution of an ionized region in an accretion disk scenario: a fully trapped H{\sc ii} region, an H{\sc ii} region trapped in the torus but not in the poles, and an H{\sc ii} region in expansion everywhere. The H{\sc ii} region would progress from one scenario to the next as its central source gets more massive and starts emitting larger amounts of ionizing radiation. We aim here to simulate this scenario and assert if and under which conditions trapped H{\sc ii} regions can be expected.

The paper is organized as follows. In Section 2 we describe the code being used and the initial conditions of our simulations. We also include in this section convergence and stability against fragmentation tests. In Section 3 we present the main results obtained from the simulations. In Section 4 we discuss how and under which conditions massive stars could form based on our results. Finally, we present our main conclusions in Section 5.

\section{Method}

The simulations are performed using the publicly available code \textsc{CMacIonize}\footnote{\url{https://github.com/bwvdnbro/CMacIonize}} \citep{BV2018}. It uses a standard finite-volume method for the hydrodynamics with an HLLC Riemann solver and is coupled to a Monte Carlo radiation transfer module, similar to the one employed in \citep{KW2004}. 

We use a static 3D cartesian grid with resolutions ranging from $64^3$ to $256^3$ cells to model an accretion torus around a star, for a range of stellar masses and ionizing luminosities. Both mass and luminosity are kept constant at any one simulation. This should not pose a problem as the timescales for the development and evolution of the HII regions in these simulations are far shorter than the timescales in which the stellar properties change.

The size of our cubic simulation box is taken to be 0.1 pc which is the same order of magnitude of the rotating cores seen around massive stars \citep{beltran}, but we can rescale the length unit of the simulation if required (see subsection \ref{subsec:HD}). The boundaries of the simulation box were set to be open so material can flow in and out of the box.
Gravity from the star is modelled through a source term in the momentum equation, as an external point mass located at the centre of the simulation box. In order to avoid numerical issues caused by the diverging gravitational force at small radii, we apply an inner spherical mask with a radius of 0.01 pc, corresponding to 10\% of our box size in length. Within this masked region we do not solve the hydrodynamics, meaning that material that flows into the mask is no longer tracked.  Boundary conditions at the mask are set to inflow, but no specific conditions are set to be satisfied at the mask radius, such that it doesn't have any dynamical effects on the rest of the simulation box.
Because we do not follow the material within the mask, we also do not account for the ionizing luminosity absorbed by the masked material. Therefore the luminosity used in each of our simulations is simply a fraction of the true luminosity emitted by the star. We do, however, compute a luminosity correction (see subsection \ref{lumcor})  which allows us to estimate the full luminosities emitted by the stars in our simulations.

We perform the simulations with an isothermal equation of state. Simulations are split in two types: hydrodynamical only (HD) simulations and simulations with both hydrodynamics and ionizing radiation (RHD). HD simulations start from a very general initial setup which is outlined in detail in subsection \ref{subsection:setup} and are let to evolve freely for 50,000 years in order to obtain a stable accretion scenario. The resulting accretion structure consists of a torus and models what we expect to find around a star that has just started emitting ionizing radiation. The torus may vary in shape and density, depending on the initial parameters. We discuss the impact of each of the input parameters has on the structure later, in the results section. Once a stable accretion structure is attained, the output of the HD simulation is used as the initial conditions for the RHD runs.

In the RHD runs, the mass at the centre also acts as a point source that emits ionizing photons isotropically. The photons can be either absorbed by the gas or re-emitted until they leave the grid \citep[for more details see][]{BV2018}. The ionizing luminosities are set based on the expected value for the mass found at the centre of the simulation box. For simplicity, throughout the paper, we use only `luminosity' to refer to the `ionizing luminosity', unless specified otherwise. At each Monte Carlo timestep we emit $10^6$ ionizing luminosity packets from the central source and iterate ten times to obtain sufficiently converged ionization fractions for each cell.  A two temperature approximation is used: fully ionized gas has a temperature $T_i = 8,000$~K whereas fully neutral gas has a temperature $T_n=500$~K, see \citet{2019KristinL}. Partially ionized gas temperatures are calculated as a linear combination of the two cases based on the actual ionization fraction of a cell.

\subsection{Initial setup}
\label{subsection:setup}

The Bondi radius, $R_B$, is defined as the radius where the accretion speed equals the sound speed for a spherically symmetric accreting star, and is given by:
 
 \begin{equation}
   R_B =  \frac{GM_\star}{2c_s^2},
 \end{equation}

 \noindent
 with $G$ being the gravitational constant, $M_\star$ the mass of the star and $c_s$ the sound speed, which itself is given by:
 
 \begin{equation}
   c_s  = \sqrt{ \frac{k T}{ \mu_m m_p}},
 \end{equation}
 where $k$ is  Boltzmann's constant, $m_p$ is the proton mass and $\mu_m$ the mean molecular weight of the gas, such that $\mu_m = 1$ for fully neutral hydrogen and $\mu_m = 0.5$ for fully ionized hydrogen. Thus, our sound speed has a specific range of values under the two temperature approximation:
 
 \begin{equation}
  \sqrt{ \frac{k T_n}{m_p}} \approx 2.0 \text{ km/s }  < c_s < 12.8 \text{ km/s }     \approx  \sqrt{ \frac{k T_i}{0.5 m_p}} 
 \end{equation}
 
 \noindent
 The values for the density and tangential velocity are set as a function of their value at the Bondi radius, $\rho_B$ and $v_B$:

 \begin{equation}
   \rho{}(r) = \rho{}_B \left(\frac{R_B}{r}\right)^{1.5}
   \label{density}
 \end{equation}
 and
 \begin{equation}
   \mathbf{v}(r) = v_B \left(\frac{R_B}{r}\right)^{0.5} \mathbf{e}_\theta{},
 \end{equation}
 
 \noindent
 Where $\mathbf{e}_\theta{}$ is the tangential unit vector in cylindrical coordinates:
 \begin{equation}
   \mathbf{e}_\theta{} = -\frac{y}{R} \mathbf{e}_x + \frac{x}{R} \mathbf{e}_y,
 \end{equation}
 with $R = \sqrt{x^2 + y^2}$ the cylindrical radius.

The Bondi density, for a Bondi accretion scenario, is related to the density at infinity through the following relation:

\begin{equation}
    \rho_B = \rho_\infty e^{3/2} \approx 4.48 \rho_\infty
    \label{equation: rhoinfty}
\end{equation}
Assuming that on scales larger than the size of our simulation box the accretion flow is Bondi-like, the density at infinity would be representative of the density of the interstellar medium.

The tangential velocity has the same radial dependence as the keplerian velocity, defined as:
\begin{equation}
     v_K = \left(\frac{GM_\star}{r}\right)^{0.5}
\end{equation}
The keplerian velocity at the Bondi radius is simply:

\begin{equation}
            v_{B} =  \left(\frac{GM_\star}{R_B}\right)^{0.5} = \sqrt{2} c_s
\end{equation}
and thus independent of the star's mass. For convenience, we start the hydrodynamical simulations with keplerian velocities at each cylindrical radius.

\subsection{Hydrodynamics}
\label{subsec:HD}
The hydrodynamic equations of conservation of mass and momentum for an isothermal equation of state, ignoring the self-gravity of the accreting gas, are:

\begin{equation}
    \frac{\partial \rho}{\partial t} + \grad( \rho \mathbf{v}) =0,
    \label{mass}
\end{equation}

\begin{equation}
\frac{\partial \mathbf{v}}{\partial t} + \mathbf{v}\cdot \grad{\mathbf{v}}  + \frac{c_s^2}{\rho}\grad(\rho) + \frac{GM_\star}{r^2} \mathbf{e}_r =0,
\label{momentum}
\end{equation}
where $\rho$, $\mathbf{v}$, $t$ and $r$ are the density, the velocity, the time and the distance respectively and $\mathbf{e}_r$ is the radial unit vector. One interesting aspect of the equations above is that they are scale invariant in density. In other words, we could multiply the density by any constant factor and this would not alter the qualitative results we get. Therefore, once we run a simulation for a star of a certain mass, we already have all the simulations for any value of the initial parameter $\rho_B$ we might have chosen.


The HD runs are also scale invariant under the ratio $M_\star/r$. Suppose we were to re-scale in space and time, such that $r$ becomes $Cr$ and such that $t$ becomes $Ct$, where $C$ is a constant re-scaling factor.
Then the mass and momentum equation become:
\begin{equation}
    C\left(\frac{\partial \rho}{\partial t} + \grad(\rho \mathbf{v} )\right) =0
    \label{mass2}
\end{equation}

\begin{equation}
C\left(\frac{\partial \mathbf{v}}{\partial t} +\mathbf{v} \cdot \grad{\mathbf{v}}  + \frac{c_s^2}{\rho}\grad(\rho) + \frac{CGM_\star}{r^2} \mathbf{e}_r \right) =0
\label{momentum2}
\end{equation}

Note that, if the mass of central star in the simulation were to be $C$ times larger ($M_\star^{new} = C M_\star^{old}$ ), then the equations would not differ from our initial ones. This implies that simulations done in a simulation box of size $L$ and stellar mass $M_\star$ will appear identical to one done with a box of size $CL$ with stellar mass $CM_\star$, with the caveat that the latter will evolve a factor of $C$ faster. 

In other words, the accretion structures formed in simulations with higher central masses have the same density and velocity profiles we would expect to see in the smaller inner regions of simulations of a lower mass star. This re-scaling can be demonstrated to work numerically by re-scaling simulations with different masses and showing that the re-scaled hydrodynamical quantities follow a common profile. This is done later in this section, in \ref{subsec:scaling}. However, it should be noted that, this is only true if the effect of self-gravity is not included. 

\subsection{Radiation hydrodynamics}
\label{subsec:RHD-setup}
In the simulations that include radiation, in addition to the hydrodynamical equations, we need to take into account the balance between photoionization and radiative recombination, which is given by the photoionizing equilibrium equation:

\begin{equation}
    Q_H(\star) = \frac{\alpha_A}{m_H^2} \int \left(1-x_H(\mathbf{x})\right)^2 \rho^2(\mathbf{x}) dV, \label{equation:ionization}
\end{equation}
where $Q_H(\star)$ is the photoionizing luminosity of the star (which for each simulation is considered a constant), $\alpha_A$ is the (constant) recombination rate to all excited states of hydrogen, $m_H$ is the mass of hydrogen and $x_H$ the neutral fraction of hydrogen and the integral is computed over the entire volume of the simulation box.

The recombination rate we use is given by the fit by \citet{Verner} which, for a temperature of 8000 K used in this work, is $\alpha_A = 4.896\times 10^{-13} {\rm cm}^3~{\rm s}^{-1}$. The diffuse radiation field is also taken into account by tracking the photons that are absorbed and re-emitted as ionizing ones. Photons recombine to the ground state at a recombination rate of $\alpha_1 = 1.773 \times 10^{-13} {\rm cm}^3{\rm s}^{-1}$ and thus the probability of a photon being re-emitted as an ionizing photon is $P_R=\alpha_1/\alpha_A \approx 0.36$.

We can obtain the volume that will be ionized for a given photoionizing luminosity by assuming a sharp transition from a completely ionized ($x_H = 0$) to a completely neutral ($x_H=1$) gas,
and by realizing that diffuse re-emission boosts the photoionizing luminosity with an additional factor $\frac{1}{1-P_R}$, such that the photoionizing equilibrium equation becomes:
\begin{equation}
   Q_H(\star) = \frac{\alpha{}_B}{m_H^2} \int_{S_0}^{S_I} \rho{}^2(\mathbf{x}) dV,
\label{eq: phot_equilibrium}
\end{equation}
where $S_0$ represents the inner boundary of the ionized volume, $S_I$ represents the ionization front (which can have an arbitrary shape) and $\alpha_B = \alpha_A(1-P_R) =  \alpha_A - \alpha_1$ is the  total radiative recombination rate to all levels above the ground state (i.e., the total radiative recombination rate to levels 2 and higher).

Here, unlike in the HD runs, a particular density has to be chosen, as this will determine the number of recombinations occurring and thus the impact a certain luminosity will have on the torus. However, the RHD simulations are scale invariant in the ratio $\rho^2/Q_H$, as the numerator determines the number of recombinations and the denominator the number of ionizations per unit time.

\subsection{Convergence}
\begin{figure}
    \centering
    \includegraphics[width=\columnwidth]{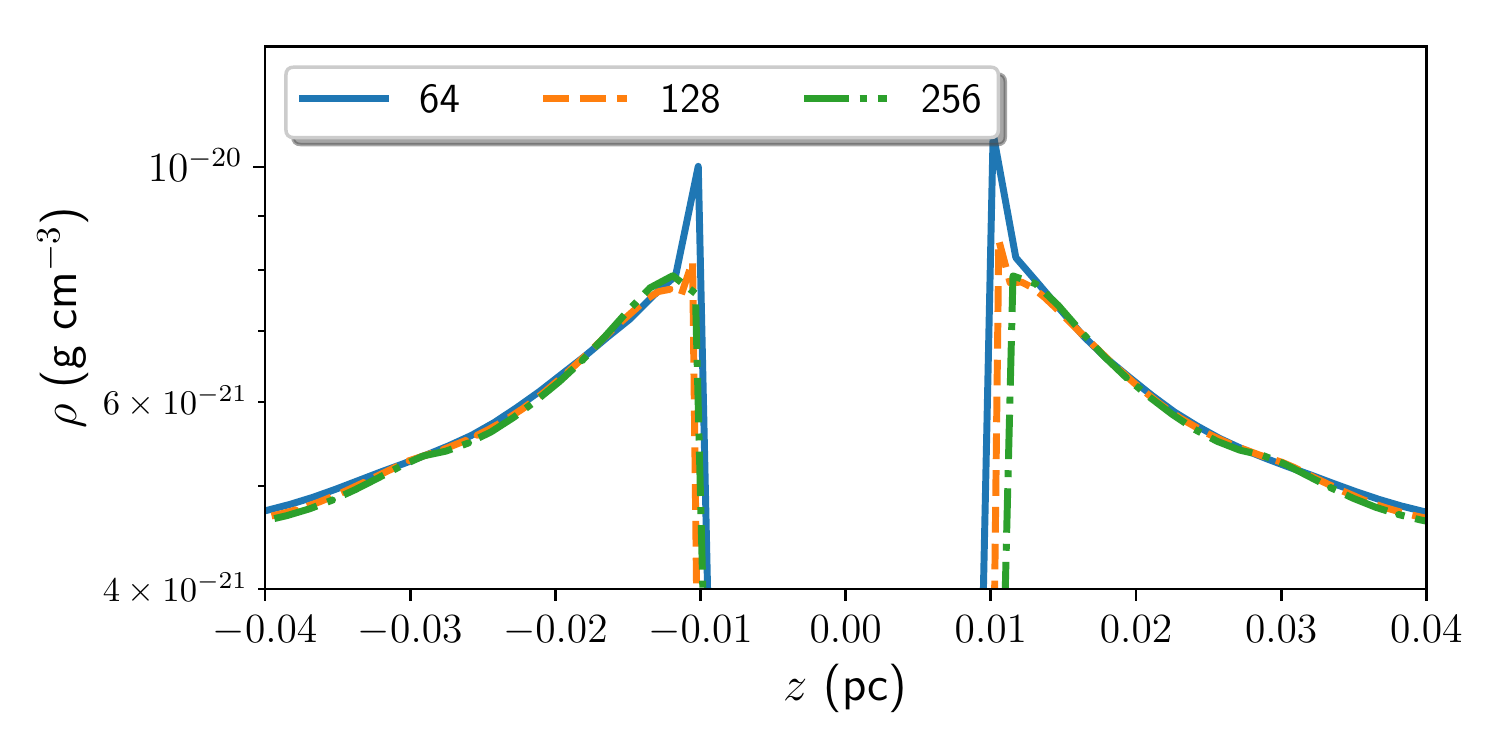}
    \caption{The density profile of for a 20 solar mass star through the torus axis for three different resolutions ($64^3$, $128^3$ and $256^3$ cells). The result for the two highest resolutions are clearly converged. The small difference at 0.01 pc is due to the fact our masked region has an integer number of cells which makes its edge slightly different for the three resolutions. }
    \label{fig: convergence}
\end{figure}

We ran the simulations at three different resolutions: $64^3$, $128^3$ and $256^3$ cells. We found that, for most of the simulations, the three resolutions present little difference when compared to one another, as can be seen in Fig.~\ref{fig: convergence}. 
The simulations in the remainder of this work have a resolution of $256^3$. We do not consider using a higher resolution as this would demand too much computational time to run all required models and would make little difference to the qualitative results obtained.

\subsection{Testing the scalability}
\label{subsec:scaling}

In section \ref{subsec:HD} we have shown analytically that the length and mass units can be re-scaled such that there is no change to the hydrodynamical evolution of the system. We explore this in this paper as a way of using one simulation to represent a number of different scenarios. We can see from Table \ref{table:scalability} that a simulation with a central mass of 100 solar masses can be re-scaled to either represent the smaller scales of a simulation with a central mass of ten solar masses or to represent larger scale of a thousand solar mass simulation.

Clearly, re-scaling implies that the units of length, mass and time have changed and, therefore, the hydrodynamical properties need to change accordingly as specified in Table \ref{table:unit_conversion}

\begin{table}
 \caption{All these simulations are equivalent according to the scaling relation. Thus we can model smaller radii within the masked regions by considering simulations of larger and larger masses} 
 \begin{tabular}{lll}
  \hline
 Scaling factor C & Central Mass ($M_\odot$) & Length \\
 \hline
 0.1 & 10  & 0.1 L\\
 1 & 100 & 1 L \\
 10 & 1000 & 10 L\\
 \hline 
 \end{tabular}
 \label{table:scalability}
\end{table} 
\begin{table}
 \caption{Re-scaling factors that need to be applied for the different hydrodynamical variables of the simulations.} 
 \begin{tabular}{lll}
 \hline
 Quantity & Units &Re-scaling \\
 \hline
 Length & [L]  & L $\longrightarrow$ CL\\
 Mass & [M]  & M $\longrightarrow$ CM\\
 Time & [T] & T $\longrightarrow$ CT \\
 Density & [M]/[L]$^3$ & $\rho \longleftrightarrow \rho/C^2$\\
 Velocity & [L]/[T] & v $\longrightarrow$ v \\
 Angular Velocity &  1/[T] & $\Omega \longrightarrow \Omega/C $\\
 \hline 
 \end{tabular}
 \label{table:unit_conversion}
\end{table}

An alternative way to see this is by investigating the units for the hydrodynamical quantities and how they link to the simulation parameters. The velocity unit immediately links to the sound speed of the gas and hence its temperature, which we assume constant for all simulations. This mean that the length unit and time unit will have a fixed ratio for all simulations. The mass unit of the system is clearly set by the choice of central point mass $M$. Once the mass and sound speed are chosen, the Bondi radius is also unequivocally fixed. In principle, this should fix the density unit, but since the solution of the hydrodynamical equations does not depend on the density (as explained in subsection \ref{subsec:HD}) this is not the case. Thus, we use a constant $\rho{}_B$, which defines the density at the Bondi radius, in order to fix the density. The only parameter left to choose is the box size, which will determine which part of the underlying general solution we sample.

As a result, we expect all simulations to have a general profile for each of the hydrodynamical quantities, which can be expressed in terms of Bondi units (Bondi radius, Bondi density, etc) and that can then be adjusted to represent any desired set of values.

\begin{figure*}
    \centering
    \includegraphics[width = \textwidth]{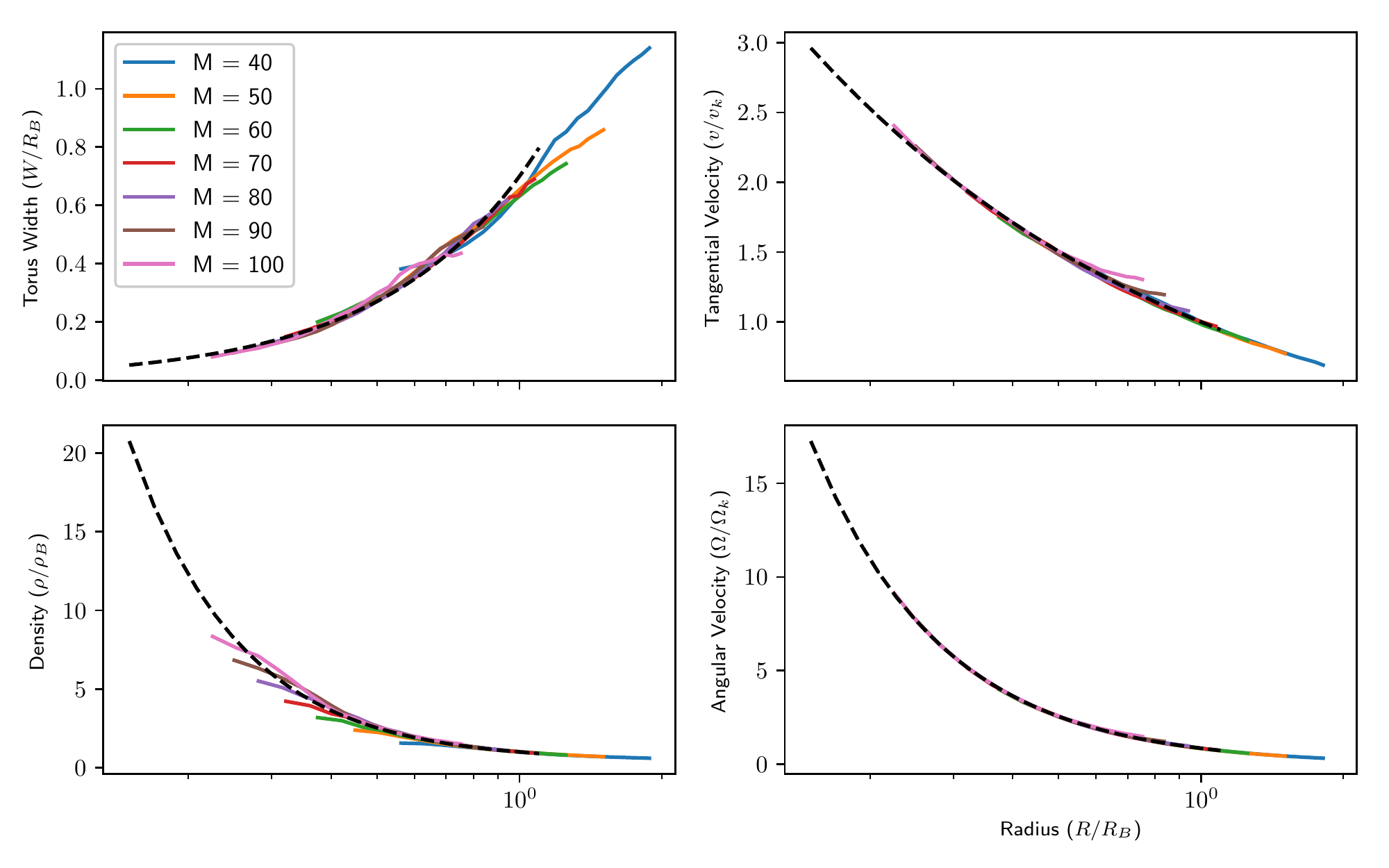}
    \caption{Basic hydrodynamical quantities re-scaled to Bondi units to show the existence of a common profile. The panels show, from top to bottom and left to right: the width of the torus, the density at the torus plane, the tangential velocity in respect to the keplerian velocity and the angular velocity in respect to the keplerian angular velocity. Every simulation had identical initial conditions except for the mass which is indicated by the colors: 40 $M_\odot$ (blue), 50 $M_\odot$ (yellow), 60 $M_\odot$ (green), 70 $M_\odot$ (red), 80 $M_\odot$ (purple), 90 $M_\odot$ (brown), 100 $M_\odot$ (pink). Power laws were fitted to each profile and are shown as dashed lines.}
    \label{fig:scaling}
\end{figure*}

In order to prove that this is indeed applicable numerically, we have re-scaled the simulations in order to show that a universal profile exists for each hydrodynamical quantity, as shown in Fig. \ref{fig:scaling} . The description of the re-scaling procedure for each quantity is laid out below.

\subsubsection{Width of the Torus}

\begin{figure}
    \includegraphics[width = 0.9 \columnwidth]{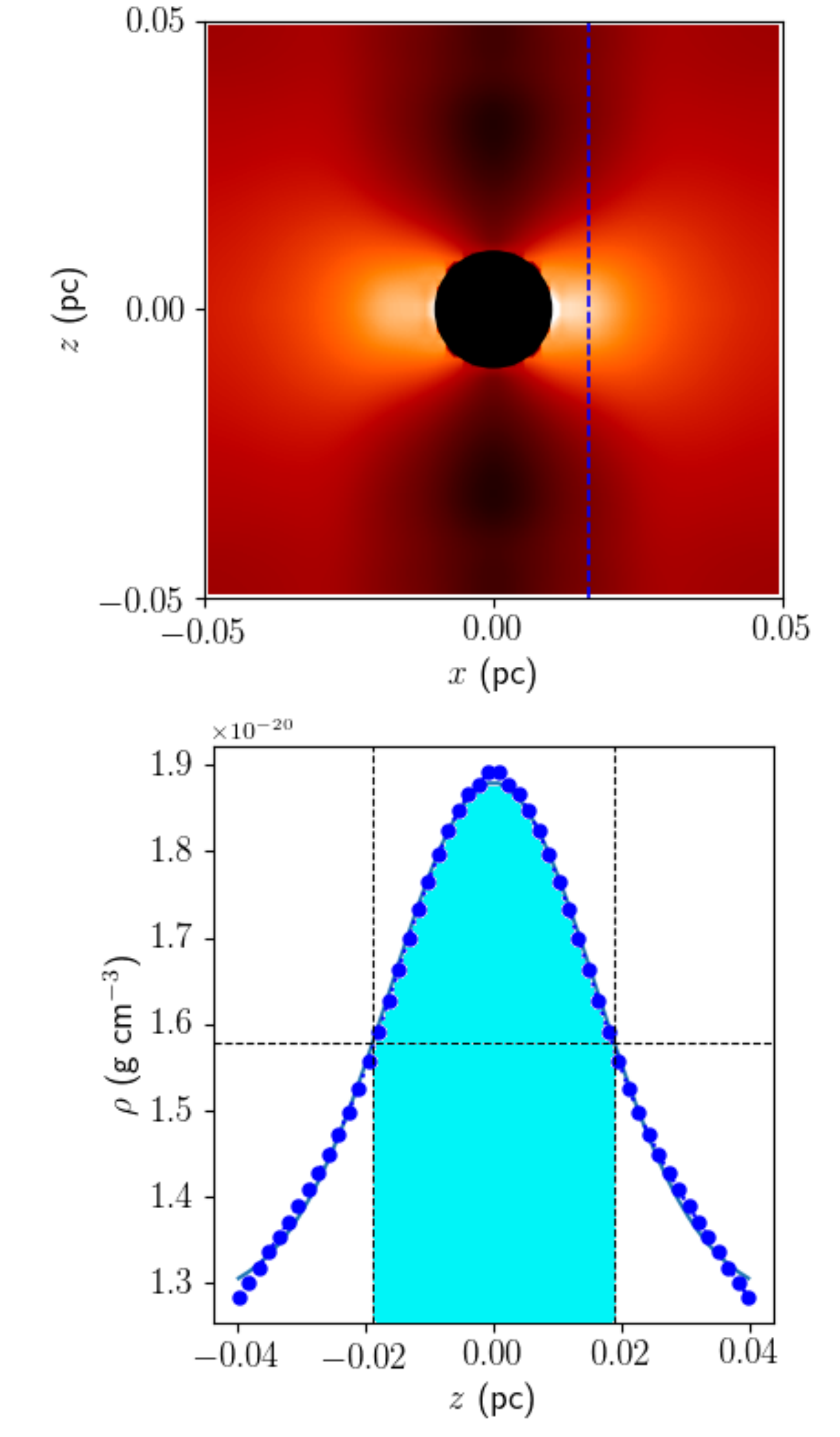}
    \caption{\emph{Top}: A density plot of a simulation of a 35 solar mass star. The blue line shows the slice used for the density profile in the bottom image. \emph{Bottom}: The density profile of the slice (blue dots) and the corresponding Gaussian fit (blue line). The dashed lines show the value for the half maximum and the heights (z) in the disk at which it occurs. The shaded light blue region shows the region that we consider to compose the torus.}
    \label{fig: gaussian_fitting}
\end{figure}

We define the torus within our simulations to have a certain thickness based on where the majority of the gas within the simulation is concentrated. In order to compute the width of the torus, we plot a density profile from a simulation slice (perpendicular to the torus plane) for each radius from the central point mass. We then fit a Gaussian to the profile and define the material that represents the torus to be the material that lies in between the half maxima points. One such density slice with the Gaussian fitting is illustrated in Fig.~\ref{fig: gaussian_fitting}.

This procedure is repeated for every simulation. These simulations are then re-scaled according to table \ref{table:scalability} and expressed in terms of Bondi units to show that their profile is the same. Finally we fit a power law $y = c\times x^{m} $ to obtain an expression for the general profile. The final fitted parameters for the power law were $m =1.51, c=  0.79$. In order to obtain the physical values for any particular simulation we can simply multiply both the radius and the width of the torus by the Bondi radius of that particular simulation.

\subsubsection{Density}

Similarly to the torus width, we first re-scale the simulations in terms of the Bondi radius. The density values are then divided by the Bondi density found for this particular re-scaling, such that $\rho_{(r = R_B)} = 1$. Recall that the density can be re-scaled by a constant factor at will. By doing that we obtain a universal profile to which we fit a power law $y = c\times x^{m} +c_0 $,  with parameters  $m =-1.88, c = 0.57, c_0 =0.43$. In order to adjust the profile to a particular simulation we multiply the radius, R, by the desired Bondi radius $R_B$ and the density by the correction factor, $\rho_{corr}$:

\begin{equation}
    \rho_{corr} = \frac{1}{M^2 \rho_B}
\end{equation}

\subsubsection{Tangential and Angular Velocity}
As discussed earlier in this section the velocity does not change with the re-scaling of the simulations, as both the time and length units are re-scaled by the same factor. The angular velocity, however, does require re-scaling as it has units of inverse time. We divide each velocity and angular velocity value by their keplerian counterpart ($v_K = \sqrt{GM/R}$ and $\Omega = v_K/R$). The values are then expressed in Bondi units such that at the Bondi radius the values for the velocity and angular velocity equal one. We then fit a power law, as it was done for the density, to the obtained profiles. The fitting values obtained for the angular velocity, $\Omega$,  $m =-1.58, c= 0.99, c_0 = 0.01$ and for the tangential velocity, $v$,  $m = -0.52, c= 1.17, c_0 = -0.17$.

The correction back to SI units is done by multiplying either simulation by their bondi counterparts:

\begin{equation}
    v_{corr} = v_B  = \sqrt{2} c_s
\end{equation}
\begin{equation}
    \Omega_{corr} = \Omega_B  = \frac{\sqrt{2} c_s}{R_B}
\end{equation}

\subsection{Toomre Q calculation}
\label{subsec:toomre}

\begin{figure}
    \centering
    \includegraphics[width= 0.5\textwidth]{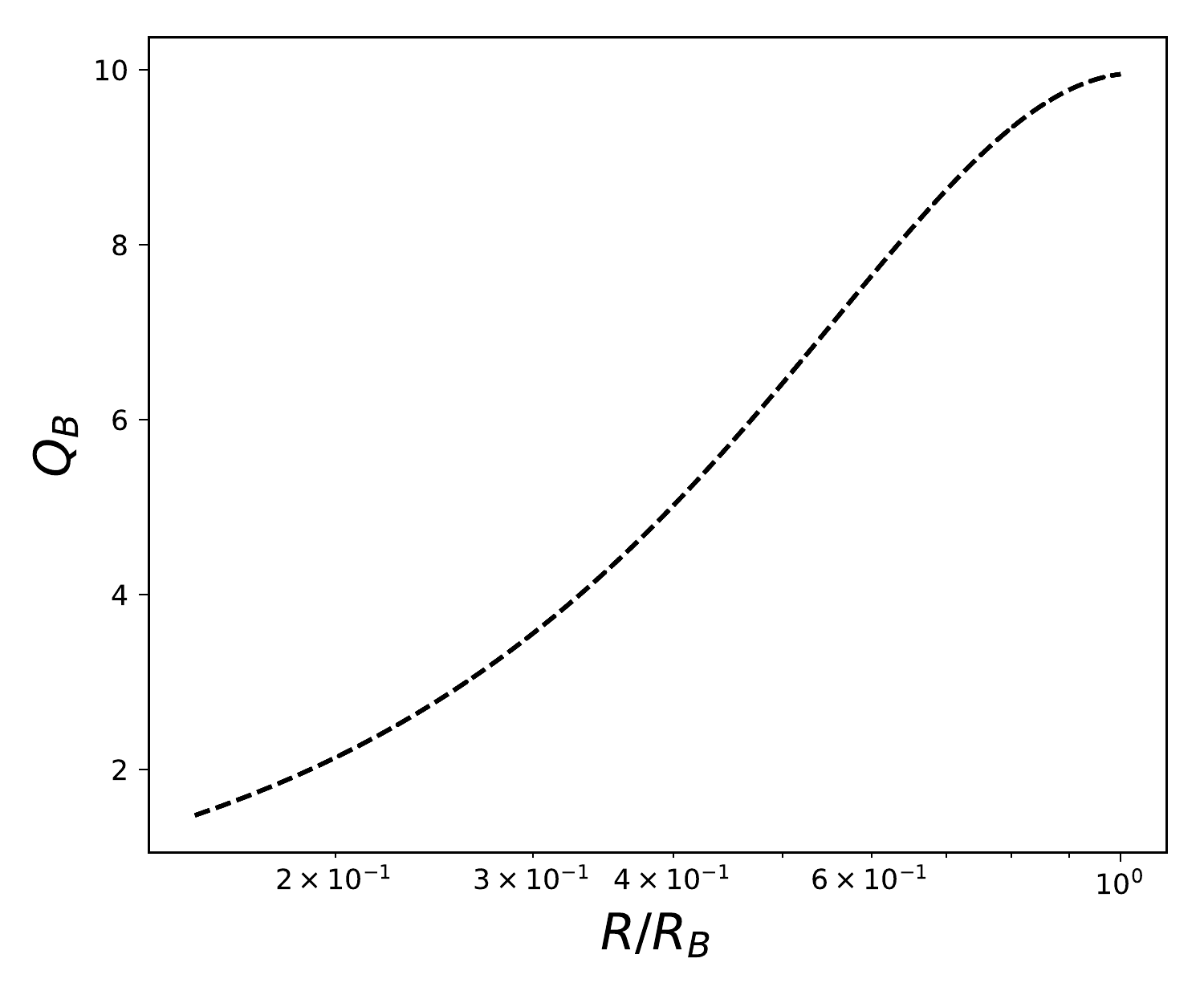}
    \caption{The profile for the $Q_B$ in Bondi units, which can be re-scaled to find the Toomre Q profile of any simulation}
    \label{fig: Toomre_Bondi}
\end{figure}

As the models we are currently studying do not include self gravity, in order to have a consistent accretion scenario we need to check if we expect fragmentation to occur within the torus. Given a thin accretion disk we can assess its local stability against fragmentation by calculating the Toomre $Q$ parameter \citep{toomre}, defined as:
\begin{equation}
    Q = \frac{c_s \kappa}{\pi G\Sigma},
\end{equation}
where $\kappa$ is the epicyclic frequency $\left(\kappa = \sqrt{r\frac{d\Omega^2}{dr} + 4\Omega^2} \right)$ and $\Sigma$ is the surface density of our disk. 
In simple terms, the numerator of the Toomre parameter is representative of the thermal ($c_s$) and rotational ($\Omega$) energies, whereas the denominator is representative of the gravitational energy.  For instability, gravity must be dominant both over the rotational and the thermal energy.
Thus, for $Q \gg 1$, the disk is stabilized by rotation and will not collapse, while for $Q \lesssim 1$ the disk is likely to fragment. Note that as we use an isothermal equation of state the value $c_s$ is fixed for the HD simulations, and, thus, the thermal support is constant with radius. The rotational support, in contrast, is larger for smaller radii where the angular velocity is higher. Therefore, the stability of the disk is directly linked to the relative rate of increase in density and velocities as we approach smaller radii, such that we have two cases:
1) if the rate of increase in angular velocity is larger than the rate of increase in density the rotational support increases more than linearly to the gravitational de-stabilizing effect, meaning that the torus would become more stable for smaller radii; 2) if the density increases more than linearly to the angular velocity, the torus gets more unstable for smaller radii.
From our universal profiles obtained in the last subsection we can see that for our simulation the second case applies and our disk will be more unstable with smaller radii. 
In order to determine if our simulations are or are not unstable, we require estimates for a disk surface density. Because the Toomre Q parameter only applies to a thin disk, for any point in the disk the scale height has to be much smaller than the radius. We use a height as 0.01 of the radius, which gives us the same error in the calculation of the Q parameter for every radius. The surface density $\Sigma = 0.01 R\times \rho$, with both R and $\rho$ being expressed in Bondi units.  We compute the ratio $Q_B =\Omega/\Sigma$ in Bondi units based on the general profiles obtained in the last section in order to obtain a universal profile for the Toomre Q parameter shown in Fig \ref{fig: Toomre_Bondi}.
The Toomre Q in Bondi units ($Q_B$) is related to the true Toomre Q of any simulation by the following expression:

\begin{equation}
    Q = \frac{c_s}{\pi G}*Q_B*F_{conv},
\end{equation}
where $F_{conv}$ is the conversion factor to SI units defined as:

\begin{align}
        F_{conv} = &  \frac{\Omega_{corr}}{\Sigma_{corr}}\\
        =&\left[\frac{\sqrt{2}c_s}{R_B}\right] \times \left[\frac{1}{M^2\rho_B R_B}\right]\\
        =&\frac{\sqrt{2}c_s}{M^2\rho_B R_B^2} 
\end{align}

\begin{figure}
    \centering
    \includegraphics[width =\columnwidth]{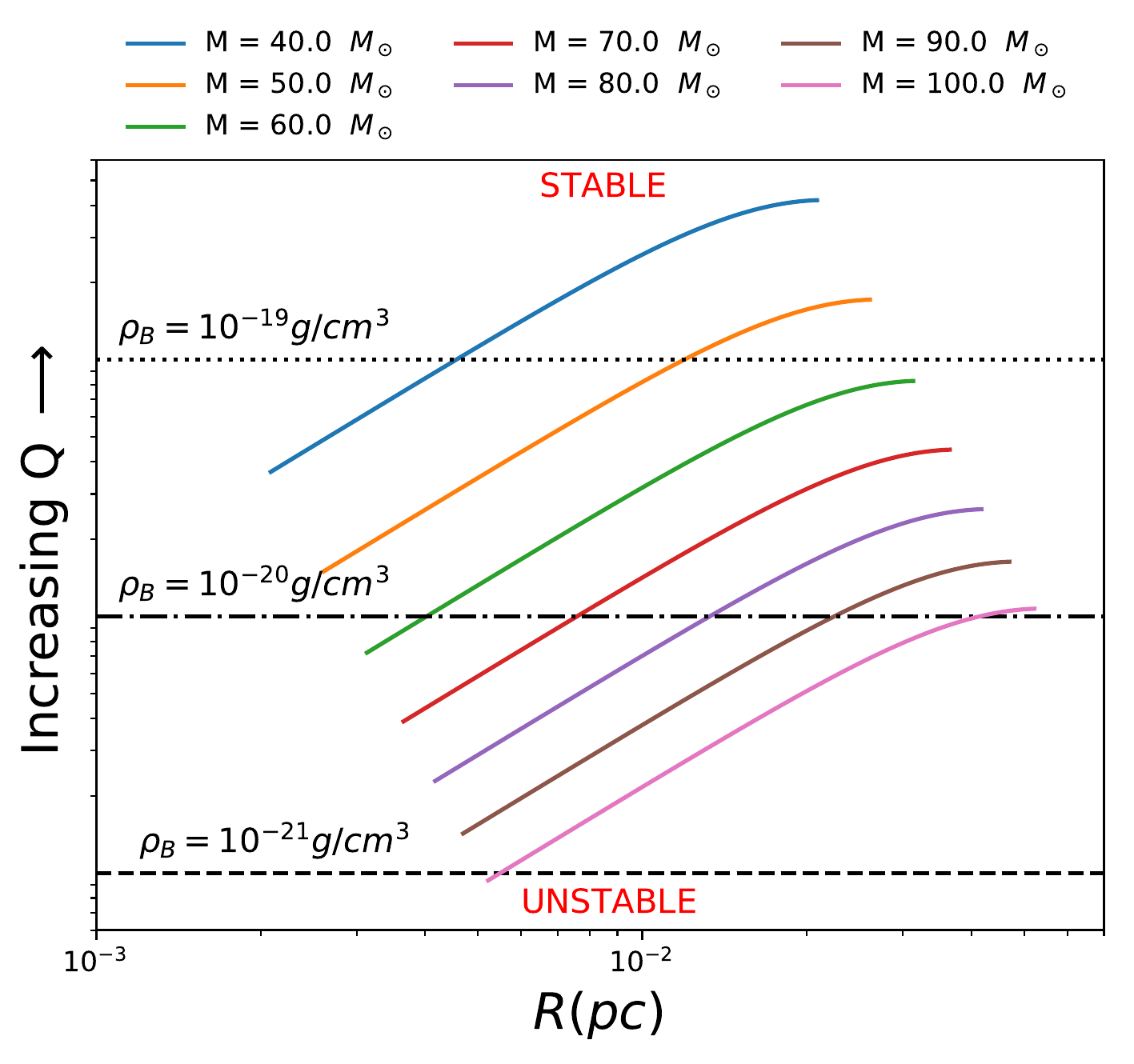}
    \caption{Toomre values for different masses. The colors associated with each mass are as in Fig. \ref{fig:scaling}. The horizontal lines show the where $Q = 1$ for different Bondi densities. Anything below a particular horizontal line is Toomre unstable and anything above is Toomre stable. }
    \label{fig: Toom_dens}
\end{figure}

\noindent
Clearly the value of $Q$ is going to differ significantly depending on which densities and masses are considered, as illustrated in Fig. \ref{fig: Toom_dens}. Nevertheless, the general trend is that the torus becomes more unstable for smaller radii.

\subsection{Ionizing Luminosity Correction}
\label{lumcor}

Our simulations use an internal mask of radius 0.01 pc. For the RHD runs, the region inside the mask is assumed to be ionized and, therefore, no radiation is absorbed. Therefore, the true luminosity of our central source is the value we set in the simulation plus an offset due to the number of photons we expect to have been absorbed inside our masked region. 

In order to have an estimate of how much radiation has been absorbed within the masked region, we make use of the homology of the hydrodynamic simulations as discussed in \ref{subsec:HD}. 
We can use the profiles obtained in Fig. \ref{fig:scaling} to model the density and and shape of the torus for any simulation for the radii smaller than the one used for our mask. Given these values we can estimate the luminosity absorbed within the mask.

The homology implies that, for a specific choice of realization with density $\rho{}(\mathbf{x})=\rho{}_B\rho{}'(\mathbf{x'})$, the ratio
\begin{equation}
    Q_{H_B} = \frac{m_H^2Q_H(\star)}{\alpha{}_B\rho{}_B^2 M_\star^4 R_B^3} = \int_{S_0'}^{S_I'} \rho'^2(\mathbf{x'}) dV',
\end{equation}
will be independent of that specific choice of realization of the general density profile, where all quantities in the integral have been replaced with their corresponding Bondi unit counterparts.
To find a general expression for the Bondi luminosity that is absorbed within some sphere with radius $S_M'$ (the radius of the mask in Bondi units), we will assume that this absorption can be reasonably approximated by absorption of the torus, with the width and density profiles following the fitted power-laws in \ref{subsec:scaling}.
Assuming cylindrical symmetry, the Bondi luminosity is then:
\begin{multline}
    Q_{H_B} = 2\pi{}\int_{r'_0}^{r'_M} \rho'^2(r') h'(r') r' dr'  = 2\pi{} \left(f({r'}_M) - f({r'}_0)\right),
    \label{Bondi_luminosity}
\end{multline}
with
\begin{align}
    f(r') &= f_0(r') + f_1(r') + f_2(r'), \\
    f_0(r') &= \frac{h_n}{h_e + 2\rho{}_e + 2} (r')^{h_e + 2\rho{}_e + 2}, \\
    f_1(r') &= \frac{2h_n\rho{}_0}{h_e + \rho{}_e + 2} (r')^{h_e + \rho{}_e + 2}, \\
    f_2(r') &= \frac{h_n \rho{}_0^2}{h_e + 2} (r')^{h_e + 2},
\end{align}

\section{Results}
\subsection{Hydrodynamical Simulations}

We first ran a number of HD simulations in order to obtain a torus to be used as initial conditions for the RHD runs.  In these runs, the gravitational contraction in directions normal to the set rotation is halted by centrifugal forces, leading to a natural flattening of the system into a torus and to a rarefaction of the polar regions.
Parameter space has been explored by changing the mass of the central star and the tangential velocity (keplerian, sub and super keplerian). Here we analyse how each of these affects the final torus structure.

\subsubsection{Changing mass}
\begin{figure}
    \centering
    \includegraphics[width = \columnwidth]{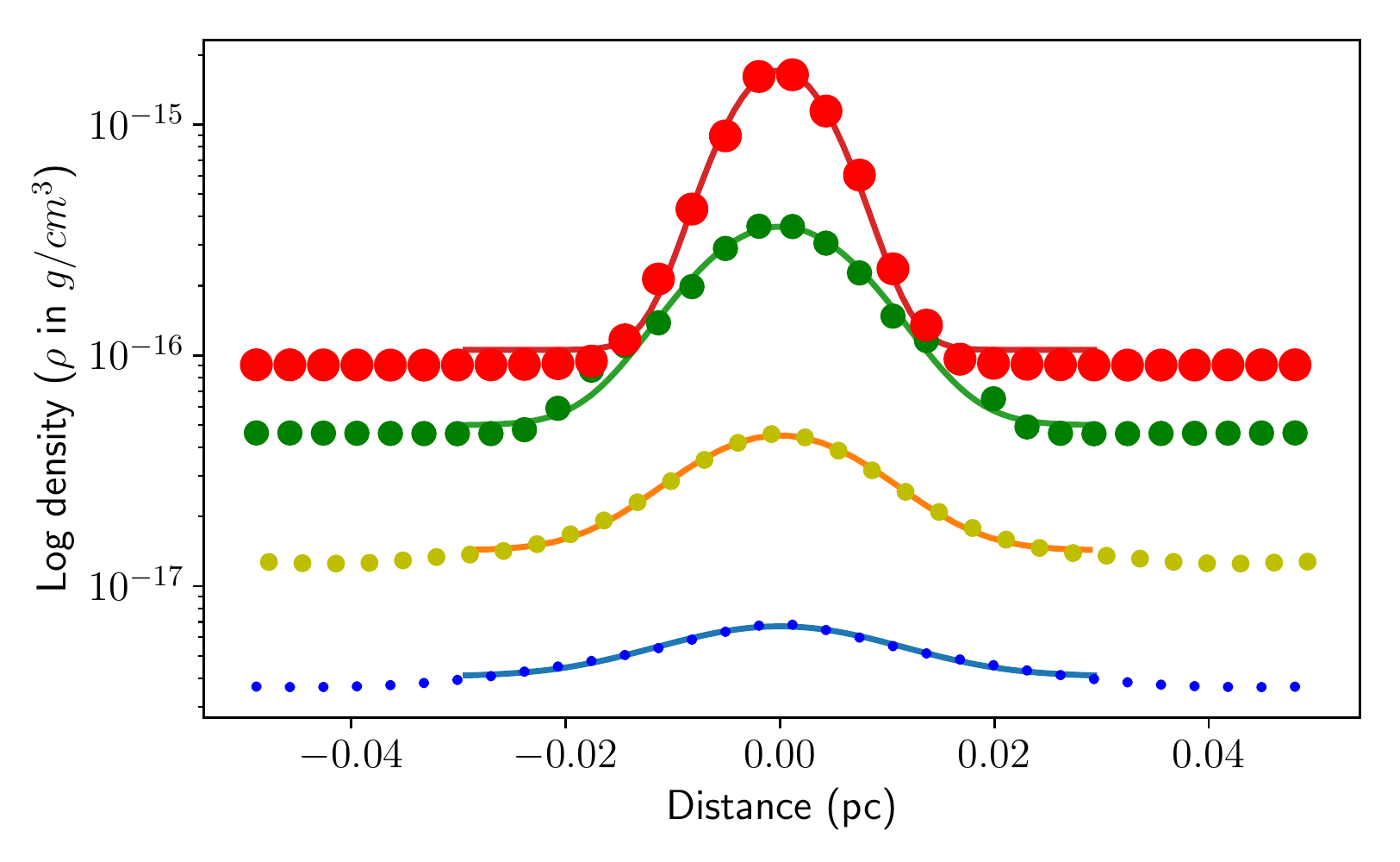}
    \caption{The Gaussian fit (as in figure \ref{fig: gaussian_fitting}) for simulations with different central masses (20, 40, 60 and 80 solar masses). The dots are weighted according to the stars mass and show the simulation data, while the lines correspond to the fitted Gaussian.  We can clearly see that as the mass increases the fitted Gaussian gets narrower and peaked at higher values indicative of a thinner, denser torus}
    \label{fig: gaussian}
\end{figure}

Due to the larger gravitational force, larger central masses lead to higher centrifugal speeds at any given radius. In turn, this leads to a stronger flattening of the torus, making the torus thinner and denser. 
This is illustrated by Fig. \ref{fig: gaussian} where we plot the Gaussian fitting used to estimate the surface density of the torus for distinct stellar masses. 

Another effect is that polar regions become more rarefied. We quantify this by the parameter  $\eta(x) =  \rho_{\rm{Torus}}(r = x)/\rho_{\rm{Polar}}(z = x)$, where $x$ is the distance from the central point mass. The value of $\rho_{\rm{Torus}}(r)$ at each x is obtained by averaging the value of the density at a certain radius in the midplane of the torus $r = x$. For the  $\rho_{\rm{Polar}}(z)$ we average the densities along the torus normal at distance $|z= x|$ from the central point mass.  We characterize the density contrast for the entire box with a single value $\eta =  \langle \eta(x) \rangle$ obtained by averaging all the values $\eta(x)$ for x values ranging from just above the mask radius to the half the box size.

\begin{figure}
    \centering
    \includegraphics[width = \columnwidth]{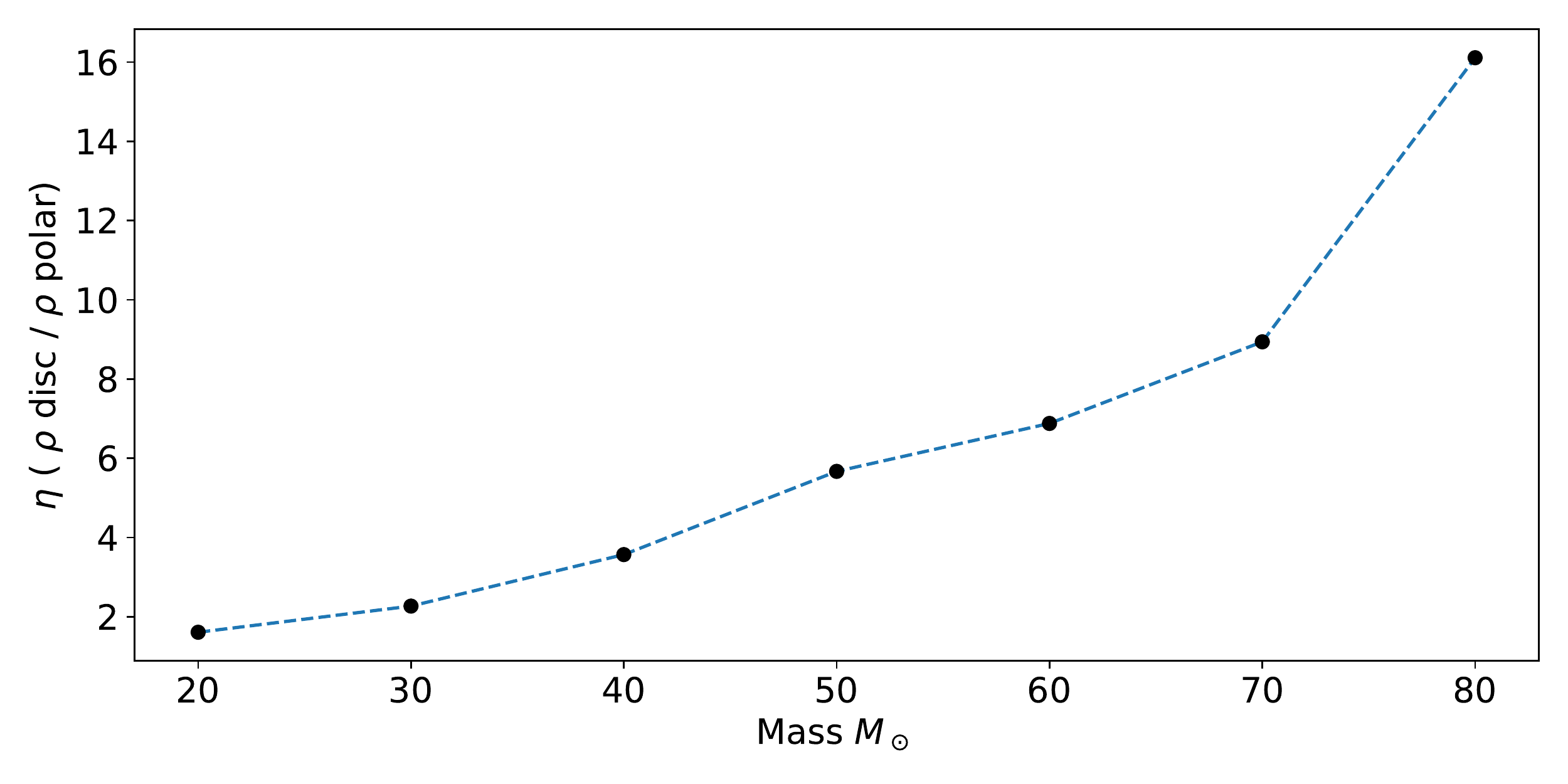}
    \caption{The evolution in the ratio of the torus and polar densities for different central masses. For lower masses the torus is only a couple of times denser than the polar regions. For higher masses the differences in density exceed an order of magnitude.}
    \label{fig: eta}
\end{figure}

The values found are shown in Fig. \ref{fig: eta}. The density contrast increases with mass and can be as low as $\eta = 2$ for a central mass of 10 solar masses and as high as $\eta = 16$ for the case with 80 solar masses. 

\subsubsection{Changing tangential velocity}

Throughout the paper we use keplerian velocities as an initial condition for our torus. Since in the HD simulations the medium is completely neutral, the temperature of the gas is 500K, making the keplerian velocity at the Bondi radius 2.8 km $\rm{s}^{-1}$. Using velocities lower than the keplerian velocity leads to a thicker, less dense torus whereas super-keplerian velocities lead to the opposite effect.

\subsection{Radiation Hydrodynamics Simulations}


 
The results of the RHD simulations can be broadly divided into three categories:

\begin{enumerate}
    \item Fully trapped H{\sc ii} region 
    \item H{\sc ii} region trapped in the torus midplane but in pressure driven expansion in the polar axis
    \item H{\sc ii} region in pressure expansion at all axis
\end{enumerate}

\begin{figure}
    \centering
    \includegraphics[width= \columnwidth]{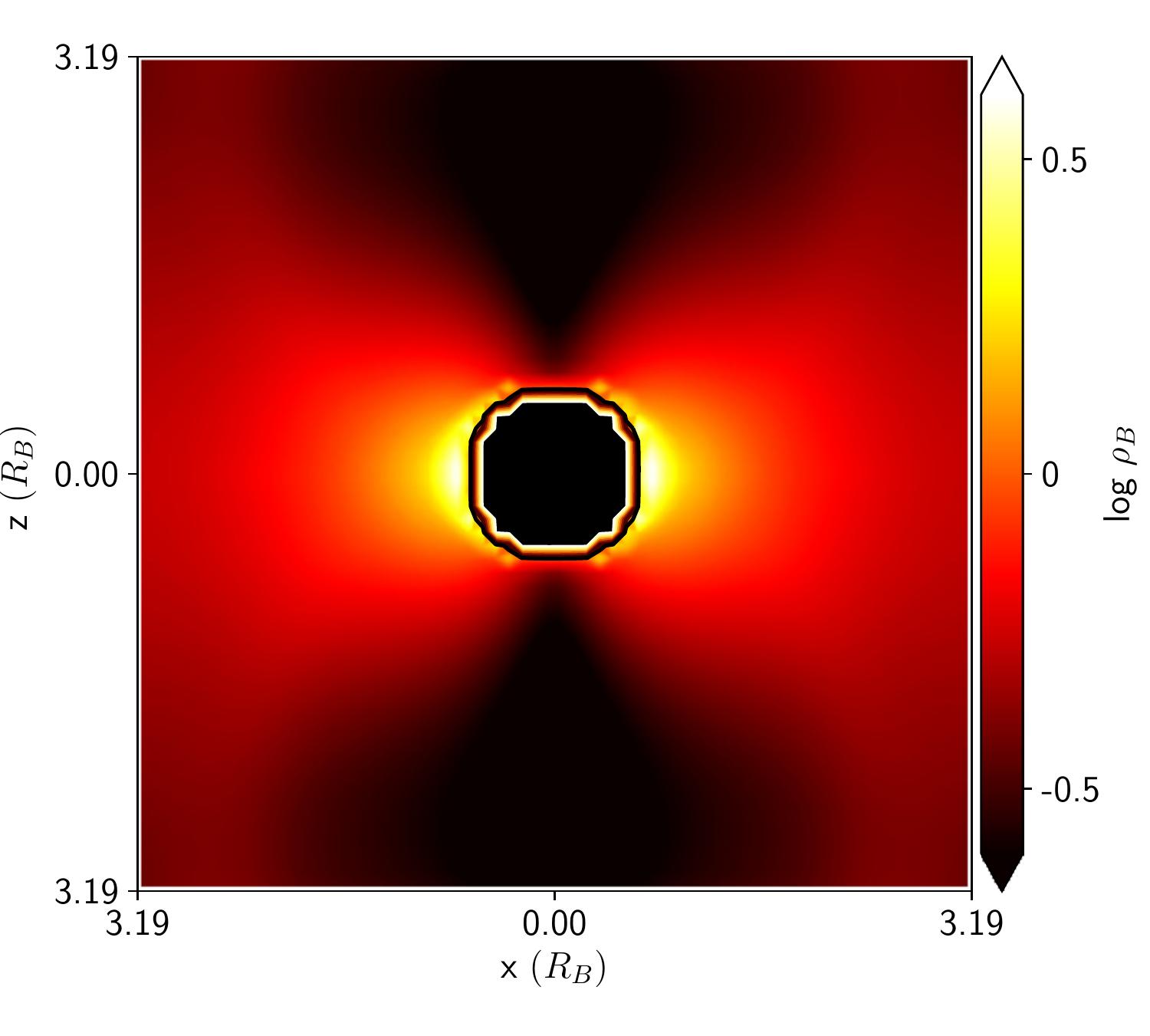}
    \caption{Density slice for a Bondi luminosity of $2.13\times 10^{-3}$. The ionization contours are shown in yellow (75\% ionized), red  (50\% ionized) and black (25\% ionized). We can see from the ionization contours that the H{\sc ii}  region is completely trapped and almost spherical in shape.}
    \label{fig: A}
\end{figure}

\begin{figure}
    \centering
    \includegraphics[width = \columnwidth]{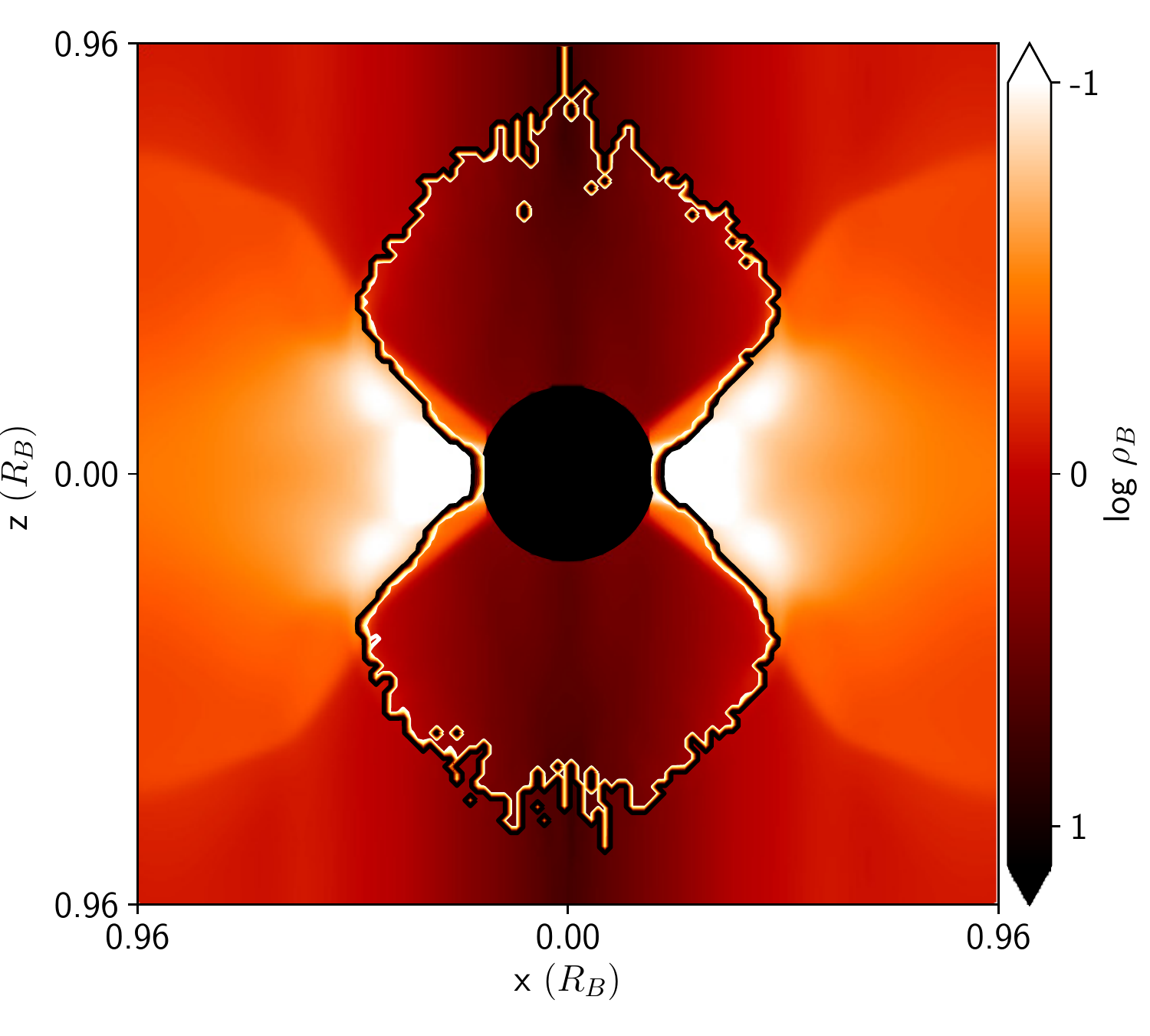}
    \caption{Density plot of a slice of an RHD simulation with a Bondi luminosity of $0.273$. The contour lines show different ionization fractions (0.75, 0.5 and 0.25 for yellow, red and black respectively). Due to the reduced density the H{\sc ii} region expands farther  in the poles than in the torus, giving the ionized region an hourglass shape. The H{\sc ii} region is not smooth due to the noise in the Monte Carlo method.}
    \label{fig: B}
\end{figure}

\begin{figure}
    \centering
    \includegraphics[width=\columnwidth]{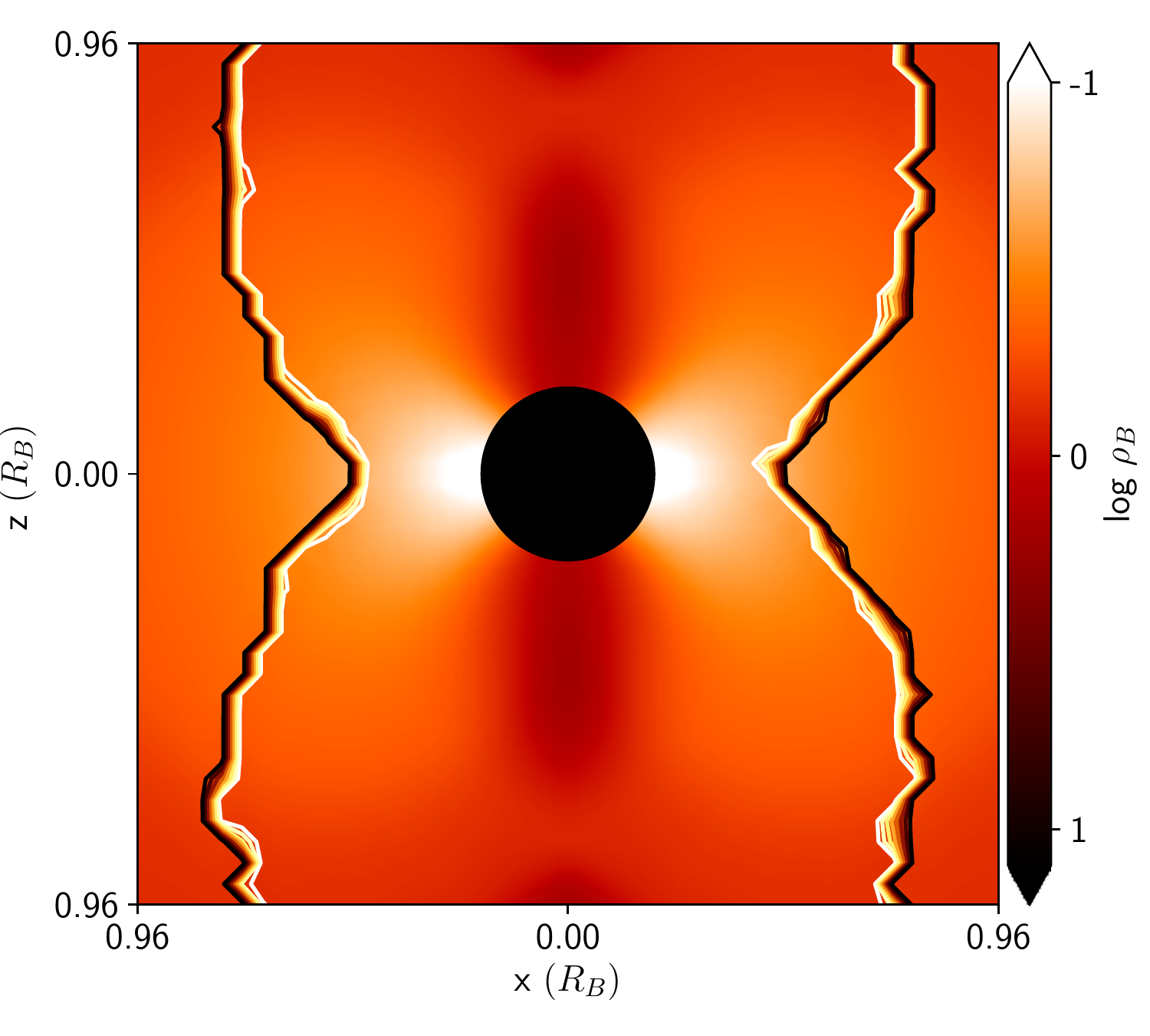}
    \caption{A 100 solar mass star simulation with luminosity of 2.73. Black, red and yellow contour lines are indicative of regions where 25\%, 50\% and 75\% of the gas is ionized, respectively. As it can be seen, the H{\sc ii} region is trapped in the torus plane but the polar regions, due to their much lower densities, have been completely ionized and have undergone pressure driven expansion.}
    \label{fig: C}
\end{figure}

\noindent
These are shown in Figs. \ref{fig: A}, \ref{fig: B} and \ref{fig: C}, respectively.
The outcome of each simulation is naturally dependant on the values of the density, which will regulate the recombination rate, and the luminosity, which will determine the number of hydrogen atoms being ionized at any point in time (see Eq.\ref{eq: phot_equilibrium}).  The simulation results will also depend on the difference in density between the torus plane and the polar axis quantified by $\eta$. If $\eta$ is close to one, there is no path through which the photons can escape more easily and a roughly spherical H{\sc ii} region forms around the star. This is seen in simulations for stars around 10 solar masses or lower (Fig.~\ref{fig: A}). 
For more massive stars the $\eta$ parameter is larger, indicating that the photons will find more resistance when travelling in the torus midplane in comparison to the polar axis. As a result the polar regions get ionized to much larger radii than the torus; leading to a roughly hourglass shaped H{\sc ii} region (see Fig.~\ref{fig: B}).
Because of this difference in density and consequently the extent of the H{\sc ii} region we can expect that the transition from a trapped to an expanding ionized region will also happen distinctly in different directions. 
If we assume, as in the spherical scenario, that transition to the expansion phase happens at some critical radius, then some parts of the H{\sc ii} region will enter the expansion phase before others.

Let the critical luminosity $Q_H^{\rm{C}}$ be the luminosity that is just large enough to ionize out to this critical radius. Owing to the fact that the density is different in different directions in our simulations, there will be a different critical luminosity for each direction. The lowest critical luminosity, $Q_H^{\rm{C}_{\rm{L}}}$, is the one that will lead to an expanding ionized region at the line of lowest density: the polar axis. Conversely, the line of highest density through the torus midplane, will be hardest to fully ionize and will have the highest critical luminosity $Q_H^{\rm{C}_{\rm{H}}}$.  Whenever the set luminosity in the simulation in between these two extremes, $Q_H^{\rm{C}_{\rm{L}}}< Q_H <Q_H^{\rm{C}_{\rm{H}}}$, the resulting scenario will be a trapped H{\sc ii} region in the torus plane and an expanding one at the poles (Fig. \ref{fig: C}).  
Because the ionizing luminosity is proportional to the density squared the range of values for which this scenario occurs also depends on the $\eta$ parameter:

\begin{equation}
    \frac{Q_H^{\rm{C}_{\rm{H}}}} {Q_H^{\rm{C}_{\rm{L}}}}\propto \left(\frac{\rho_{\rm{torus}}}{\rho_{\rm{Polar}}}\right)^2 \sim \eta^2
\end{equation}
From Fig. \ref{fig: eta} we can see that a larger stellar mass implies a larger value of $\eta$. As a result, the luminosity range that leads to case (ii) will increase with mass too.

\begin{figure*}
    \centering
    \includegraphics[width=2\columnwidth]{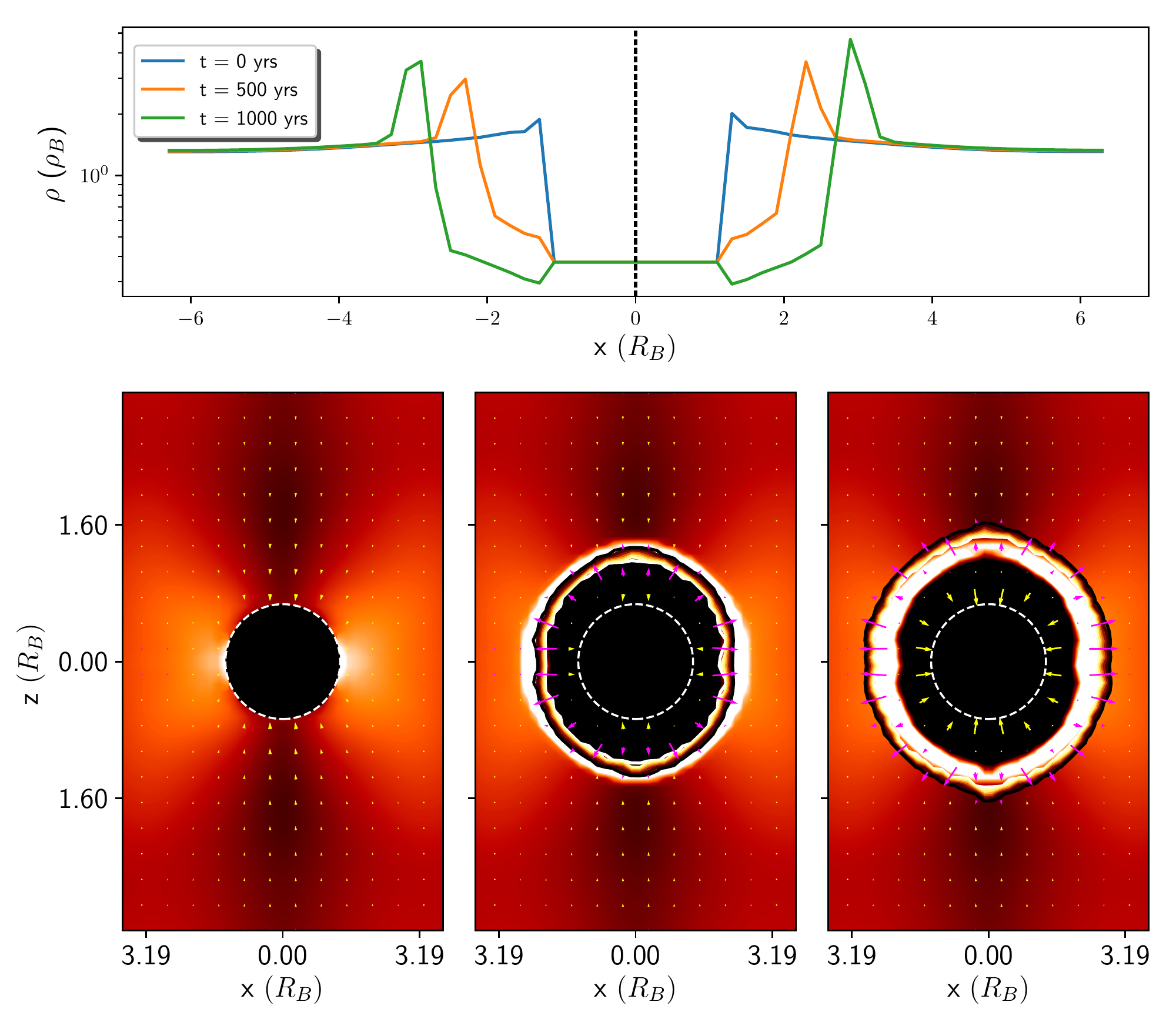}
    \caption{\emph{Top}: Torus density profiles at different times showing the evolution of an expanding H{\sc ii} region with Bondi luminosity $2.73$. 
    \emph{Bottom}: Density slices of the simulation in the top panel. Degrees of ionization are shown by the black, red and yellow lines as in Fig \ref{fig: C}. Pink and yellow arrows show outflows and inflows respectively.  The whole H{\sc ii} region is ionized past the critical radius and enters the expansion phase. In the last snapshot we can observe that the inside of the H{\sc ii} region is drained as material inside the ionized region accretes to the star, but no more material traverses the ionization front.
}
    \label{case3}
\end{figure*}

Finally, there is the case where the luminosity is large enough to ionize both polar and torus axis out to a radius larger than their critical radii ($Q_H > Q_H^{\rm{C}_{\rm{H}}}$), creating an expanding ionized bubble (Fig. \ref{case3}).
One could, therefore, think of our three possible scenarios presented above as a natural evolution of the system. As the mass and, consequently, the ionizing luminosity increase, an initially fully trapped H{\sc ii} region could get progressively larger, until it enters the expansion phase at the poles. The star would stop accreting through the poles but accretion would continue through the torus plane. Luminosity will continue rising until the star emits enough luminosity to ionize the torus beyond the critical radius. This would halt accretion completely, setting the final mass of the star.

\section{Discussion}
As it was said in the introduction, massive stars need to be able to continue accreting after the onset of fusion, which leads to the emission of copious amounts of ionizing radiation. It is therefore essential that H{\sc ii} regions remain trapped for timescales long enough to allow the required mass build up to occur. In respect to the factors under consideration in our simulation, the ability to trap the H{\sc ii} region is going to depend on the following:
\begin{itemize}
    \item Critical radius -- set by the value of the central point mass 
    \item Ionizing Luminosity -- set by the source(s) at the centre
    \item Torus Density -- set by the initial density and velocity and by the central point mass
\end{itemize}
In the rest of the section we discuss how the trapping of the H{\sc ii} region might occur in light of our results as well as discussing limitations of our approach to the problem.
\subsection{Luminosity Corrections}
The results presented assume that the ionizing radiation incoming from the mask is isotropically emitted from its centre. Furthermore, the luminosity escaping the mask must be only a fraction of the one emitted by the star. We can, however, correct our luminosities using the method described in \ref{lumcor} which is based on the profiles obtained for the torus width and the density. 

\subsubsection{Testing the correction}

\begin{figure}
    \centering
    \includegraphics[width = \columnwidth]{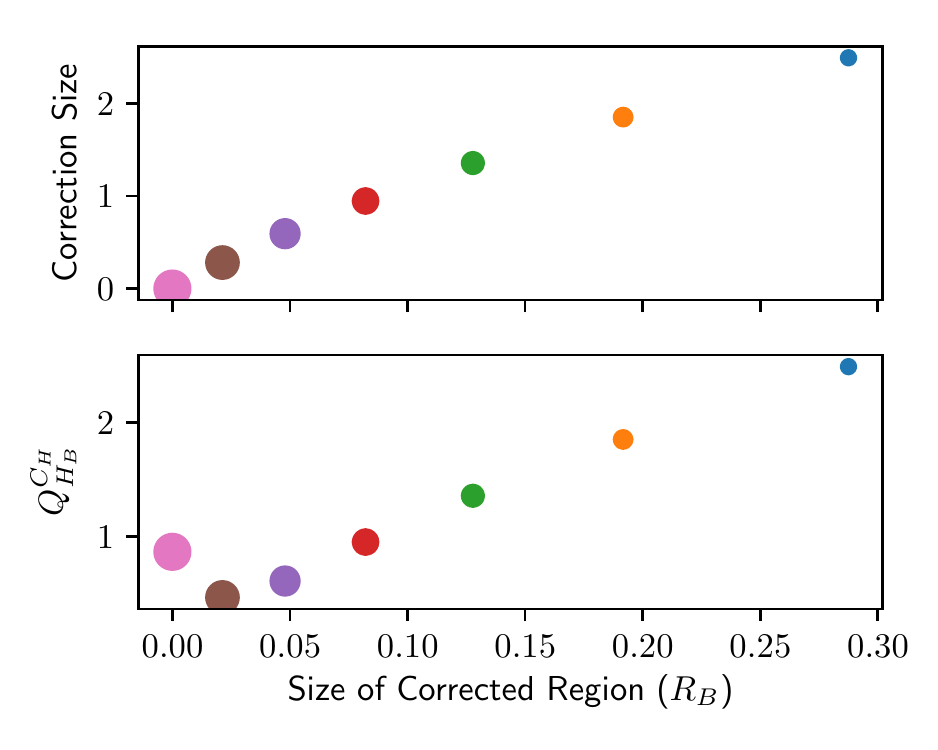}
    \caption{\emph{Top}: Correction in Bondi luminosites for different masses as a function of the size of the corrected region (equivalent to the difference in radii of the re-scaled mask for the simulation in question and that of the 100 solar mass simulation). Colors are the same as in Figs \ref{fig:scaling} and \ref{fig: Toom_dens} the and dot's size is scaled according to the mass it represents.  By definition the correction to the 100 solar mass simulation is zero. \emph{Bottom}: Critical Bondi luminosities for the 100 solar mass (uncorrected) and lower mass simulations (corrected to include radiation absorption from radii above the mask of the 100 solar mass simulation)}
    \label{fig:Correction}
\end{figure}

In order to test this luminosity correction we first perform a mock correction to a known value. The scalability implies that for rescaled simulations, the size of the mask is inversely proportional to the mass. Thus, we may verify if we can reproduce the critical luminosity values obtained by a high mass ($M_{large}$) simulation by correcting the luminosity absorbed between its rescaled mask radius and the rescaled masked radius of an otherwise identical lower mass simulation, $M_{small}$:

\begin{equation}
    Q_{HB}^{C_H}(M_{large}) =  Q_{HB}^{C_H}(M_{small}) + 2\pi{}\int_{R_{Mask}(M_{large})}^{R_{Mask}(M_{small})} \rho^2 h r dr 
\end{equation}
We use the the critical luminosity for expansion in the midplane $Q_{HB}^{C_H}$ for the $100M_\odot$ as a proxy and compare it to the critical values of simulations from 40 to 90 solar masses corrected for the radiation absorbed in the region between the re-scaled masked radii. The values we use for the critical luminosity are the lowest luminosity value which caused the midplane of the disk to enter pressure driven expansion for a simulation of a particular mass. The original values (before re-scaling) are laid out in table \ref{tab: crit_lum} for reference, all for simulations with $\rho_B = 1\times10^{-20}$.
\begin{table}
 \caption{Critical luminosities for different masses. These values are not corrected to account for luminosity absorbed within the mask.}
   \begin{tabular}{lll}
    \hline
    \parbox[t]{2cm}{ Mass of \\ Central Star ($M_\odot$) } & \parbox[t]{2cm}{Log of Critical \\ Luminosity ($s^{-1}$)} \\[3ex]
    \hline
    40    & 44\\
    50    & 44.5\\
    60    & 45\\
    70    & 46\\
    80    & 47\\
    90    & 48.5\\
    100    & 50\\
    \hline 
    \end{tabular}
    \label{tab: crit_lum}
\end{table}
The test correction is illustrated in Fig. \ref{fig:Correction}. The top figure shows the relative size of the correction between the different simulations in Bondi units. The bottom plot shows the corrected value obtained for the critical luminosities. Should our values of the critical luminosity and corrections be exact, all critical luminosities should be the same. However, with the exception of the 90 and 80 solar mass simulations, the critical values obtained are larger than the expected value. This indicates that our correction provides an upper limit for the critical values. In other words, simulations for which the ionized region in the torus is in pressure driven expansion for our corrected simulations would also be in pressure driven expansion for a simulation run with a much smaller mask.

Hence, even though we cannot claim the sizes of the H{\sc ii} region would be the same as the ones in our simulations, we are able to find an upper limit in the transition between the trapped and expanding scenarios within the torus midplane. Therefore, we can obtain an upper limit on the maximum luminosity a star can have while still being able to sustain an accretion disk from which it can keep accreting mass.

\subsubsection{Luminosity comparison to models}
 We are interested in analysing which scenarios would create a suitable environment for a massive star to continue accreting via a disk given the large amount of ionizing radiation it emits. With that aim we correct the critical luminosity values obtained in our simulations to a radius of $1\times 10^{-4}$pc and compare these values with ionizing luminosities given by stellar evolution models. This radius for the correction was chosen because all our simulations, except one, are unstable at this radius, such that all calculations of luminosities are overestimates of the true value. Here we use estimates of the true ionizing luminosities of massive stars which were obtained from the stellar evolution model of  \citet{2003Sternberg}. The values for the critical luminosities for different densities (dashed lines) plotted against model luminosities (continuous green line) are found in Fig. \ref{fig: lum_comp}. For Bondi densities  of $ \rho_B =10^{-19} \rm{g}~\rm{cm}^{-3}$ and larger, we can expect the H{\sc ii} region to always be trapped in the midplane. For Bondi densities of $ \rho_B =10^{-21} \rm{g}~\rm{cm}^{-3}$ and smaller we can expect the H{\sc ii} region to be  in expansion. For intermediate Bondi densities, $10^{-21} \leq \rho_B \leq 10^{-19} \rm{g}~\rm{cm}^{-3}$, the H{\sc ii} region starts expanding within the midplane for some particular mass.

\subsection{Stability via multiplicity}

\begin{figure}
    \centering
    \includegraphics[width = \columnwidth]{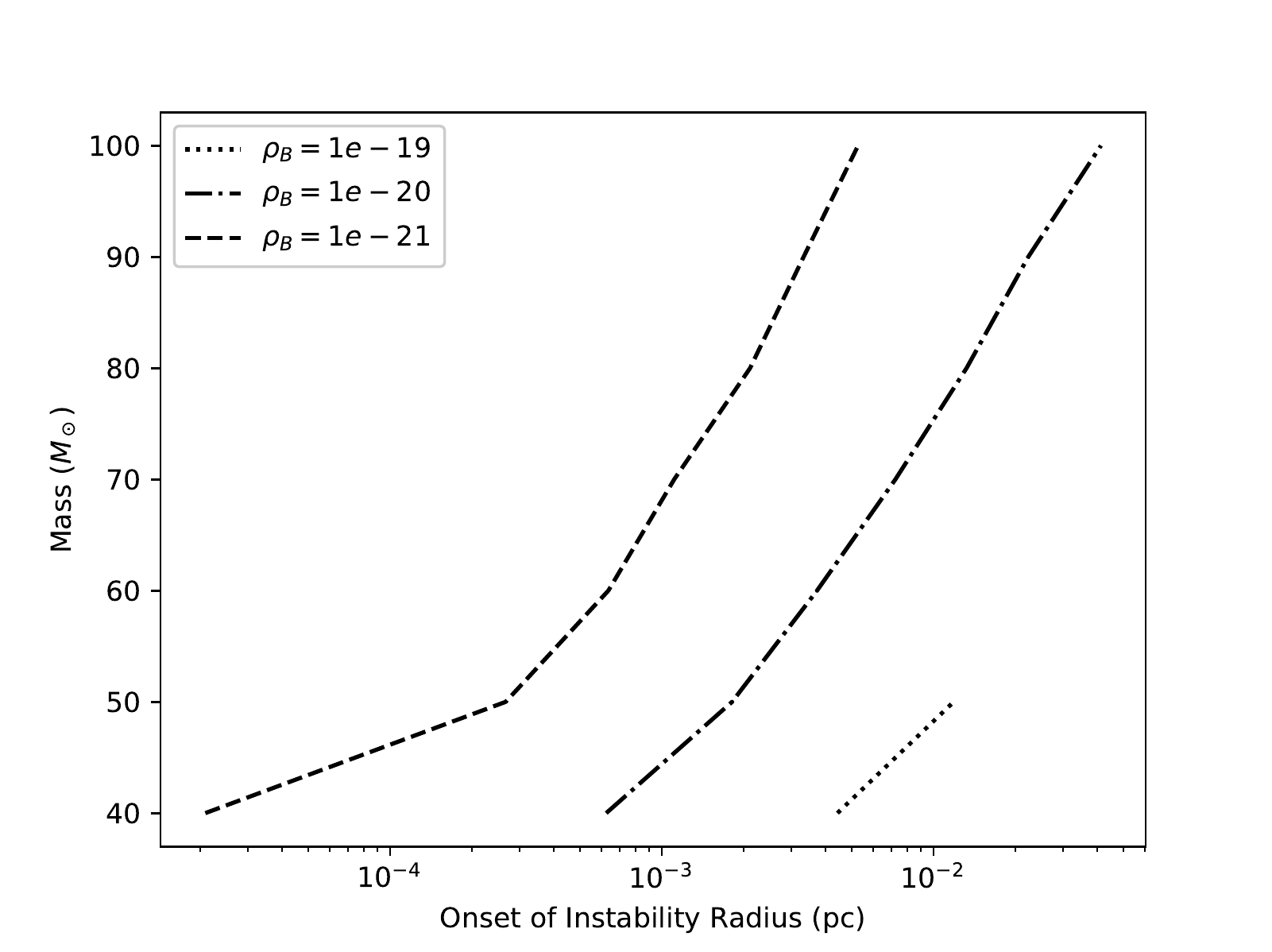}
    \caption{The radii bellow which the simulations would be Toomre unstable plotted for different stellar masses and Bondi densities. The curve for the $\rho_B = 10^{-19}$ only has values for masses between 40 and 50 solar masses because for all other masses it is completely unstable for every radius, as shown in Fig \ref{fig: Toomre_Bondi}.}
    \label{fig: q=1}
\end{figure}

As explained in the last subsection, the fate of the torus and H{\sc ii} region is strongly dependant on the density. One aspect that might be of importance to the analysis of the disk resistance against the ionizing radiation is the stability of the disk itself. In Fig. \ref{fig: q=1} we plot the radius at which we expect fragmentation to start taking place which is based on the data already shown in Fig \ref{fig: Toom_dens}. We can see that for a number of scenarios studied we expect fragmentation to occur.

 \begin{figure}
    \centering
    \includegraphics[width = \columnwidth]{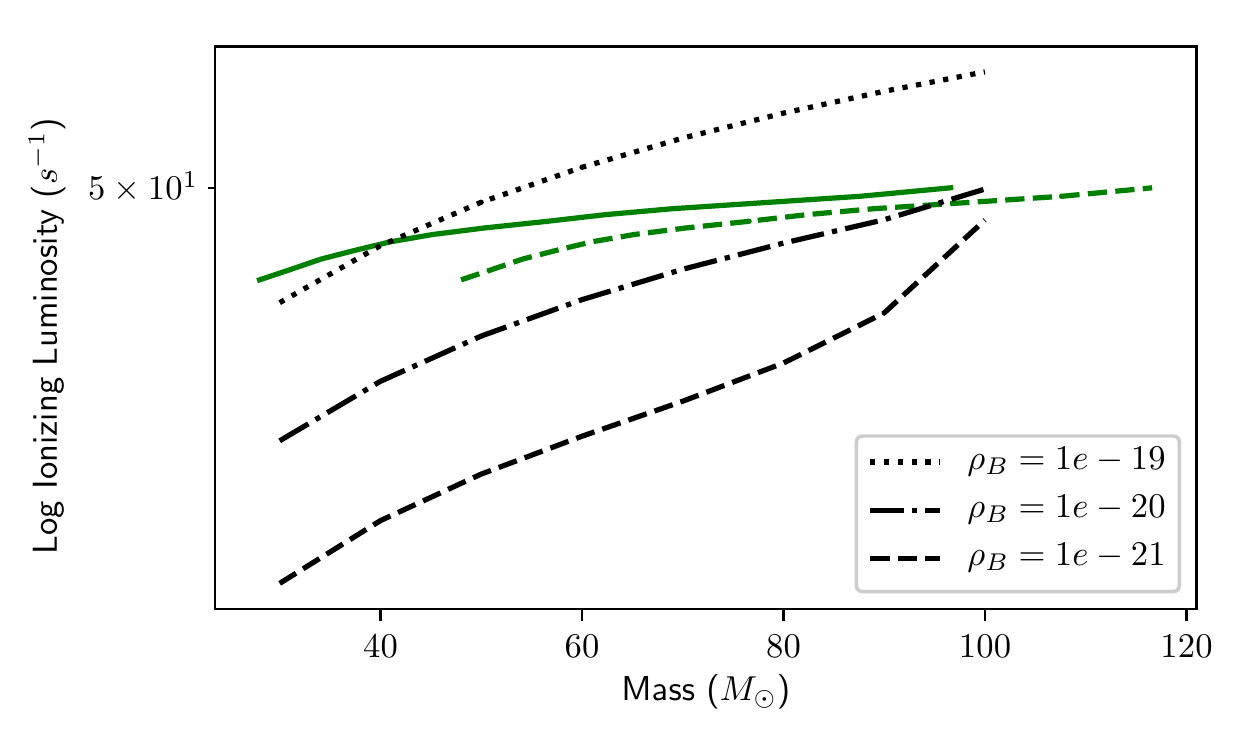}
    \caption{Dashed lines: critical luminosities for trapping H{\sc ii} regions in the torus plane for different Bondi densities. Continuous green line: the expected luminosities for a each stellar mass according to the stellar model of \citet{2003Sternberg}. If the expected values are above the critical luminosity the H{\sc ii} region will enter the pressure expansion phase in the disk axis. Dashed green line: mimics the effect of having a massive star emitting at the expected values but also having companion stars with a combined mass of 20 solar masses. We can see that for this line the number of cases for which the H{\sc ii} region is trapped increases}
    \label{fig: lum_comp}
\end{figure}

We speculate on the impact such fragmentation could have on our results. One of the effects fragmentation could have is on the interaction of radiation and the torus. Fragmentation implies that clumpy regions would form and, alongside them, lower density channels through which luminosity can escape \citep{2013Wood}. This means that regions at larger radii in the torus might get more easily ionized than our simulations would suggest. In turn this might mean that the torus potentially enters D-type expansion for considerably lower luminosities than the upper limit found. 
Fragmentation could also lead to the formation of other proto-stars in the inner region of the torus. Interestingly, if other smaller proto-stars form, they could contribute to the gravitational potential more than they contribute to the increase in ionizing luminosity. As discussed previously, a larger gravitational attraction would imply that the torus would evolve to be denser and thinner and therefore more difficult to ionize.
In order to model this effect we plot the model luminosities as if they were emitted by a star 20 solar masses larger than the true value. This mimics a massive star accompanied by a number of proto-stars with combined mass of 20 solar masses, but which do not emit significant amounts of ionizing radiation in comparison to the massive star. This is represented in Fig. \ref{fig: lum_comp} by the dashed green line. As it can be seen, the increase in mass has a stabilizing effect: the masses that would enter the expansion phase for the $ 10^{-19} \rm{g}~\rm{cm}^{-3}$ curve (portion for which the green line - true luminosity - is above the dashed line - critical luminosity values) are now all trapped (the green line is now fully bellow the dashed line). For the $ 10^{-20} \rm{g}~\rm{cm}^{-3}$ curve this also implies that the trapped scenario would occur for lower masses than if we had a single star. 
Therefore if more than one star is forming simultaneously to a large mass star the torus/disk would be more stable against ionizing radiation and could keep feeding the stars for longer. 
This is interesting as O stars are found to be in a binary and multiple system much more often than other types of stars \citep{Sana}. Perhaps, this high binarity is linked to the fact that maintaining a torus to form a massive star is more likely if there is core multiplicity. In this case, massive stars are not necessarily only competing for material, but actually need the other companion stars to form. 
However a verification if this is actually the case would require simulations with self-gravity that properly analyse the effect of fragmentation of the inner disk and its interaction with the incoming ionizing radiation which is beyond the scope of this paper and is deferred to a later study.

\subsection{Torus evolution}

\begin{figure}
    \centering
    \includegraphics[width =\columnwidth]{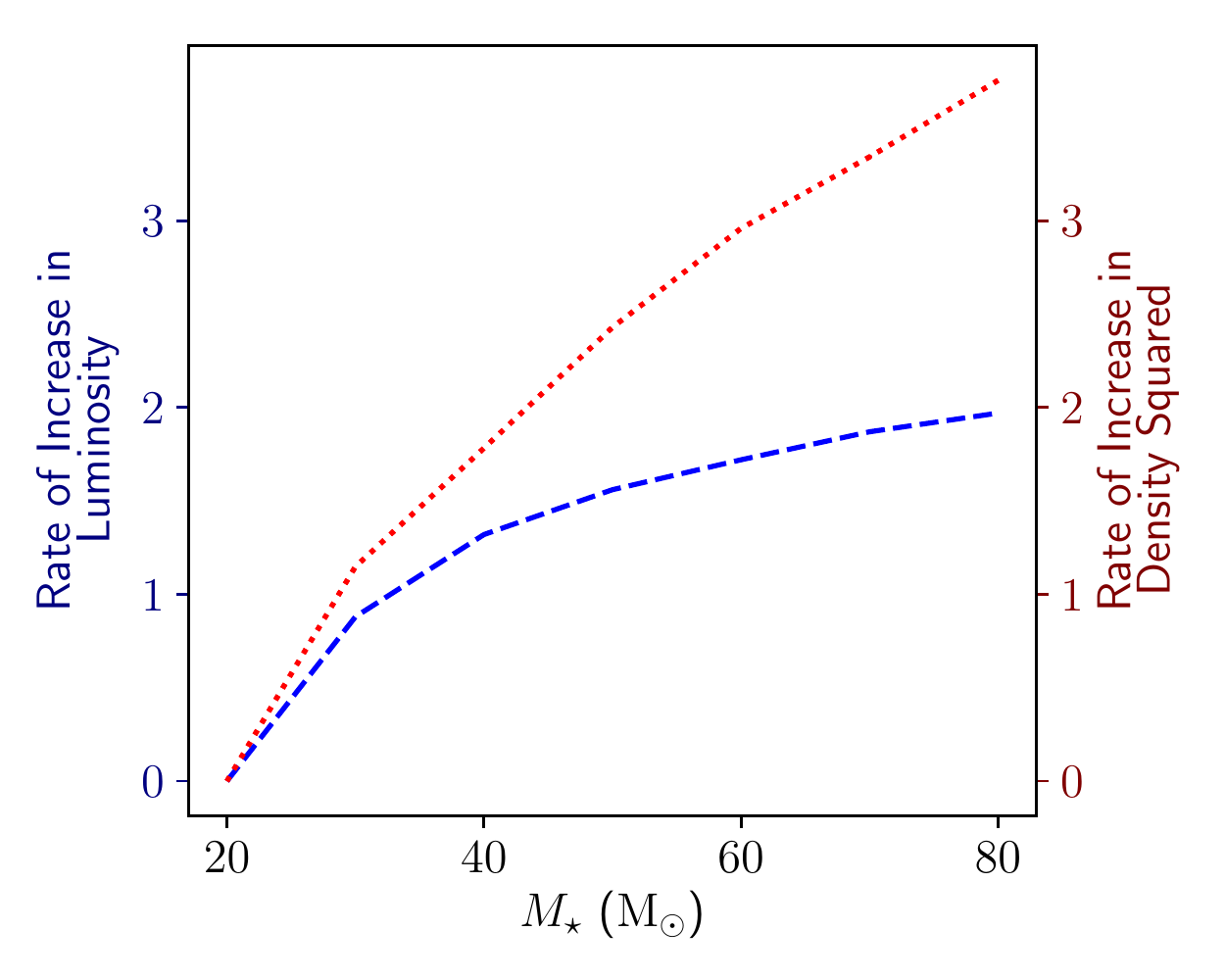}
    \caption{
The rate of increase in torus density and ionizing luminosity as a star increases its mass. The values are normalized with respect to the torus density and ionizing luminosities of a 20 solar mass star.
}
    \label{fig: lumvsdens}
\end{figure}

Although we do not simulate a self consistent evolution of the mass of our star and the corresponding torus changes it is important that we understand how the trapping of the H{\sc ii} region would change as a star gets more massive. It was pointed out previously that for higher masses the torus in our simulations is found to be thinner and denser. As a star accretes, it is therefore reasonable to assume that its torus will try to adapt to the star's mass gain by following a similar trend. Clearly it takes time for the torus material to adapt to a new configuration dictated by the central mass. If the time taken by the torus to readjust is much longer than the accretion timescale the torus will not change dramatically its shape or density throughout the star's accretion history. Hence, it is likely the torus of the massive stars is going to be less dense and easier to ionize than what we consider in our simulations.  If, conversely, the torus is able to conform quickly to the growth of the star then our simulations for different masses for a given Bondi density, can be thought of as snapshots through the evolution of a single star. We can use this extreme scenario as an upper limit for how much the torus can be expected to change throughout the stellar evolution. Fig \ref{fig: lumvsdens} shows the relative increase in ionizing luminosity (based on the stellar evolution model by \citet{2003Sternberg} ) and the torus surface density squared for this limiting case. 

Recall that the size of the H{\sc ii} region depends on the ionization balance and that the amount of ionizing photons required scales with the square of the density.  We can see from  Fig \ref{fig: lumvsdens} that the rate of increase in density squared on the torus axis is much larger than the rate of increase in luminosity of the star as it evolves. We can conclude that, if a star were to evolve in such a way, once an H{\sc ii} region is trapped in the torus it will either always be trapped, or it will quench as the luminosity coming from the star will be able to ionize up to progressively smaller radii in the torus plane. In this case the star's final mass needs to be halted by other means than ionizing feedback. 

\subsection{Mass reservoir}

Until now we have assumed that the mass of the torus and the star could keep increasing indefinitely, that is, we assumed an infinite mass reservoir. A more likely scenario would be to have a finite amount of mass that can be accreted.
Material located at a distance $R_{lim}$  from the central star for which the free-fall time is larger than the period for which the massive star accretes will clearly never be able to contribute to the star's mass. Therefore, it follows straightforwardly that the mass of the star will never be able to exceed the mass available within the boundaries of  $R_{lim}$.
In addition, limiting the amount of material available can limit the maximum density of the torus thus making the torus optically thinner and more prone to having an expanding H{\sc ii} region.

\subsection{Mask Effects}

As pointed out in section 2 we use an inner mask of 0.01 pc around our point mass. It is, therefore, reasonable to question how the results might change had we considered a smaller mask. 
It was already shown earlier in this section that the luminosity values corrected to include radiation absorbed in most of the masked region give a good estimate for the H{\sc ii} region scenario (expanding or trapped) within the torus. The size of the mask does imply that for the the H{\sc ii} region trapped within the first few cells from the mask could be more compact (trapped at smaller  radii in the disk). It is however not the aim of this paper to quantify the extent of the ionized region, but instead to obtain a general view of possible scenarios and constrain expected luminosities for which the ionized regions would no longer be trapped.

It is worth pointing out that any simulation of a torus or a disk has a masked region in the sense that accretion onto the stellar surface is not resolved. Even though some simulation resolve accretion to a single cell, this cell is still equivalent to a very small masked region in which the hydrodynamics and radiative transfer effects cannot be taken into account. The size of the mask used in our simulations was based on the fact that, at mask scales and below, a number of simulations \citep{2010Peters, 2016Klassen} as well as observations \citep{2017Beuther} indicate the existence of substructure and sometimes multiple accreting stars \citep{2010Peters, 2007K}. In light of the lack of self gravity in our simulation and, henceforth, our inability to evaluate such scales appropriately we have set a mask around these smaller scales and focused on the larger picture effect of ionizing radiation on the environment.
In addition, should a disk truly exist around a single star at scales smaller than the mask used, we can still re-scale the simulations to these smaller scales. In other words, the qualitative results regarding the possible shapes of the H{\sc ii} regions would still hold, albeit for different luminosities and density values.

\section{Conclusions and outlook}

Massive stars must extend their accretion timescale by accreting through trapped H{\sc ii} region.  We find three different scenarios for H{\sc ii} regions:
\begin{enumerate}
    \item Trapped in both torus and polar regions
    \item Trapped in the torus and in D-type expansion in the polar region
    \item In D type expansion in all directions.
\end{enumerate}

We show we expect trapped H{\sc ii} regions for simulations with initial density at the Bondi radius of  $ \rho_B =10^{-19} \rm{g}~\rm{cm}^{-3}$ or higher. Regardless of the central stellar mass if an H{\sc ii} region is trapped at one point, the ionized region will remain trapped as the star gains more mass, unless some process other than ionization is responsible for a change in the torus structure.
The simulations show that if the central mass is increased the torus structure is denser and more resistant to radiation given a certain luminosity. This indicates that if multiple stars are forming alongside the massive star such that they contribute to the gravitational potential but not to the emission of ionizing luminosity it would be possible to sustain the existence of disk for longer.

More simulations are needed including self-gravity to have a thorough understanding of how ionization and disks interact in cases where fragmentation is expected to take place. This particular topic is going to be addressed in future work.
Other factors may play an essential role in the stability and structure of the torus, such as magnetic fields, stellar winds and radiation pressure. 
Future simulations will explore these effects looking at each in turn.
Although our model is idealized, it allows us to focus our investigation on the impact of photo-ionization on the torus which would have been challenging to attribute specifically to ionization in a more complete model. 

\section*{Acknowledgements}
NS would like to thank CAPES for a graduate research funding. BV and KW acknowledge support from STFC grant ST/M001296/1. DFG thanks the Brazilian agencies CNPq (no. 311128/2017-3) and FAPESP (no. 2013/10559-5) for financial support.

We would like also to thank Ian Bonnell for helpful advice.




\bibliographystyle{mnras}
\bibliography{ref}

\begin{thebibliography}{}
\makeatletter
\relax
\def\mn@urlcharsother{\let\do\@makeother \do\$\do\&\do\#\do\^\do\_\do\%\do\~}
\def\mn@doi{\begingroup\mn@urlcharsother \@ifnextchar [ {\mn@doi@}
  {\mn@doi@[]}}
\def\mn@doi@[#1]#2{\def\@tempa{#1}\ifx\@tempa\@empty \href
  {http://dx.doi.org/#2} {doi:#2}\else \href {http://dx.doi.org/#2} {#1}\fi
  \endgroup}
\def\mn@eprint#1#2{\mn@eprint@#1:#2::\@nil}
\def\mn@eprint@arXiv#1{\href {http://arxiv.org/abs/#1} {{\tt arXiv:#1}}}
\def\mn@eprint@dblp#1{\href {http://dblp.uni-trier.de/rec/bibtex/#1.xml}
  {dblp:#1}}
\def\mn@eprint@#1:#2:#3:#4\@nil{\def\@tempa {#1}\def\@tempb {#2}\def\@tempc
  {#3}\ifx \@tempc \@empty \let \@tempc \@tempb \let \@tempb \@tempa \fi \ifx
  \@tempb \@empty \def\@tempb {arXiv}\fi \@ifundefined
  {mn@eprint@\@tempb}{\@tempb:\@tempc}{\expandafter \expandafter \csname
  mn@eprint@\@tempb\endcsname \expandafter{\@tempc}}}

\bibitem[\protect\citeauthoryear{{Beltr{\'a}n} \& {de Wit}}{{Beltr{\'a}n} \&
  {de Wit}}{2016}]{beltran}
{Beltr{\'a}n} M.~T.,  {de Wit} W.~J.,  2016, \mn@doi [\aapr]
  {10.1007/s00159-015-0089-z}, \href
  {http://adsabs.harvard.edu/abs/2016A%26ARv..24....6B} {24, 6}

\bibitem[\protect\citeauthoryear{{Beuther}, {Walsh}, {Johnston}, {Henning},
  {Kuiper}, {Longmore}  \& {Walmsley}}{{Beuther} et~al.}{2017}]{2017Beuther}
{Beuther} H.,  {Walsh} A.~J.,  {Johnston} K.~G.,  {Henning} T.,  {Kuiper} R.,
  {Longmore} S.~N.,   {Walmsley} C.~M.,  2017, \mn@doi [\aap]
  {10.1051/0004-6361/201630126}, \href
  {http://adsabs.harvard.edu/abs/2017A%26A...603A..10B} {603, A10}

\bibitem[\protect\citeauthoryear{{Bonnell} \& {Bate}}{{Bonnell} \&
  {Bate}}{2006}]{Bonnell.and.Bate.2006}
{Bonnell} I.~A.,  {Bate} M.~R.,  2006, \mn@doi [\mnras]
  {10.1111/j.1365-2966.2006.10495.x}, \href
  {http://adsabs.harvard.edu/abs/2006MNRAS.370..488B} {370, 488}

\bibitem[\protect\citeauthoryear{{Bonnell}, {Bate}, {Clarke}  \&
  {Pringle}}{{Bonnell} et~al.}{2001}]{2001.Bonnell}
{Bonnell} I.~A.,  {Bate} M.~R.,  {Clarke} C.~J.,   {Pringle} J.~E.,  2001,
  \mn@doi [\mnras] {10.1046/j.1365-8711.2001.04270.x}, \href
  {http://adsabs.harvard.edu/abs/2001MNRAS.323..785B} {323, 785}

\bibitem[\protect\citeauthoryear{{Cesaroni}, {Galli}, {Lodato}, {Walmsley}  \&
  {Zhang}}{{Cesaroni} et~al.}{2007}]{cesaroni}
{Cesaroni} R.,  {Galli} D.,  {Lodato} G.,  {Walmsley} C.~M.,   {Zhang} Q.,
  2007, Protostars and Planets V, \href
  {http://adsabs.harvard.edu/abs/2007prpl.conf..197C} {pp 197--212}

\bibitem[\protect\citeauthoryear{{Haemmerl{\'e}}, {Eggenberger}, {Meynet},
  {Maeder}  \& {Charbonnel}}{{Haemmerl{\'e}} et~al.}{2016}]{Haemmerle}
{Haemmerl{\'e}} L.,  {Eggenberger} P.,  {Meynet} G.,  {Maeder} A.,
  {Charbonnel} C.,  2016, \mn@doi [\aap] {10.1051/0004-6361/201527202}, \href
  {http://adsabs.harvard.edu/abs/2016A%26A...585A..65H} {585, A65}

\bibitem[\protect\citeauthoryear{{Harries}, {Douglas}  \& {Ali}}{{Harries}
  et~al.}{2017}]{2017harries}
{Harries} T.~J.,  {Douglas} T.~A.,   {Ali} A.,  2017, \mn@doi [\mnras]
  {10.1093/mnras/stx1490}, \href
  {http://adsabs.harvard.edu/abs/2017MNRAS.471.4111H} {471, 4111}

\bibitem[\protect\citeauthoryear{{Jankovic} et~al.,}{{Jankovic}
  et~al.}{2019}]{jankovic}
{Jankovic} M.~R.,  et~al., 2019, \mn@doi [\mnras] {10.1093/mnras/sty3038},
  \href {http://adsabs.harvard.edu/abs/2019MNRAS.482.4673J} {482, 4673}

\bibitem[\protect\citeauthoryear{{Kahn}}{{Kahn}}{1974}]{kahn}
{Kahn} F.~D.,  1974, \aap, \href
  {http://adsabs.harvard.edu/abs/1974A%26A....37..149K} {37, 149}

\bibitem[\protect\citeauthoryear{{Kennicutt}}{{Kennicutt}}{1998}]{Kennicutt}
{Kennicutt} Jr. R.~C.,  1998, \mn@doi [\araa] {10.1146/annurev.astro.36.1.189},
  \href {http://adsabs.harvard.edu/abs/1998ARA%26A..36..189K} {36, 189}

\bibitem[\protect\citeauthoryear{{Keto}}{{Keto}}{2002}]{Keto_trapped1}
{Keto} E.,  2002, \mn@doi [\apj] {10.1086/343794}, \href
  {http://adsabs.harvard.edu/abs/2002ApJ...580..980K} {580, 980}

\bibitem[\protect\citeauthoryear{{Keto}}{{Keto}}{2003}]{2003.keto.spherical}
{Keto} E.,  2003, \mn@doi [\apj] {10.1086/379545}, \href
  {http://adsabs.harvard.edu/abs/2003ApJ...599.1196K} {599, 1196}

\bibitem[\protect\citeauthoryear{{Keto}}{{Keto}}{2007}]{2007Keto}
{Keto} E.,  2007, \mn@doi [\apj] {10.1086/520320}, \href
  {http://adsabs.harvard.edu/abs/2007ApJ...666..976K} {666, 976}

\bibitem[\protect\citeauthoryear{{Klassen}, {Pudritz}, {Kuiper}, {Peters}  \&
  {Banerjee}}{{Klassen} et~al.}{2016}]{2016Klassen}
{Klassen} M.,  {Pudritz} R.~E.,  {Kuiper} R.,  {Peters} T.,   {Banerjee} R.,
  2016, \mn@doi [\apj] {10.3847/0004-637X/823/1/28}, \href
  {http://adsabs.harvard.edu/abs/2016ApJ...823...28K} {823, 28}

\bibitem[\protect\citeauthoryear{{Krumholz}, {Klein}  \& {McKee}}{{Krumholz}
  et~al.}{2007}]{2007K}
{Krumholz} M.~R.,  {Klein} R.~I.,   {McKee} C.~F.,  2007, \mn@doi [\apj]
  {10.1086/519305}, \href {http://adsabs.harvard.edu/abs/2007ApJ...665..478K}
  {665, 478}

\bibitem[\protect\citeauthoryear{{Krumholz}, {Klein}, {McKee}, {Offner}  \&
  {Cunningham}}{{Krumholz} et~al.}{2009}]{2009Krumholz}
{Krumholz} M.~R.,  {Klein} R.~I.,  {McKee} C.~F.,  {Offner} S.~S.~R.,
  {Cunningham} A.~J.,  2009, \mn@doi [Science] {10.1126/science.1165857}, \href
  {http://adsabs.harvard.edu/abs/2009Sci...323..754K} {323, 754}

\bibitem[\protect\citeauthoryear{{Kuiper} \& {Hosokawa}}{{Kuiper} \&
  {Hosokawa}}{2018}]{2018.kuiper}
{Kuiper} R.,  {Hosokawa} T.,  2018, \mn@doi [\aap]
  {10.1051/0004-6361/201832638}, \href
  {http://adsabs.harvard.edu/abs/2018A%26A...616A.101K} {616, A101}

\bibitem[\protect\citeauthoryear{{Lund}, {Wood}, {Falceta-Gon{\c c}alves},
  {Vandenbroucke}, {Sartorio}, {Bonnell}, {Johnston}  \& {Keto}}{{Lund}
  et~al.}{2019}]{2019KristinL}
{Lund} K.,  {Wood} K.,  {Falceta-Gon{\c c}alves} D.,  {Vandenbroucke} B.,
  {Sartorio} N.~S.,  {Bonnell} I.~A.,  {Johnston} K.~G.,   {Keto} E.,  2019,
  \mn@doi [\mnras] {10.1093/mnras/stz621}, \href
  {http://adsabs.harvard.edu/abs/2019MNRAS.tmp..609L} {}

\bibitem[\protect\citeauthoryear{{Massey}}{{Massey}}{2003}]{Massey}
{Massey} P.,  2003, \mn@doi [\araa] {10.1146/annurev.astro.41.071601.170033},
  \href {http://adsabs.harvard.edu/abs/2003ARA%26A..41...15M} {41, 15}

\bibitem[\protect\citeauthoryear{{McKee} \& {Tan}}{{McKee} \&
  {Tan}}{2003}]{turbulent.core}
{McKee} C.~F.,  {Tan} J.~C.,  2003, \mn@doi [\apj] {10.1086/346149}, \href
  {http://adsabs.harvard.edu/abs/2003ApJ...585..850M} {585, 850}

\bibitem[\protect\citeauthoryear{{Mestel}}{{Mestel}}{1954}]{Mestel}
{Mestel} L.,  1954, \mn@doi [\mnras] {10.1093/mnras/114.4.437}, \href
  {http://adsabs.harvard.edu/abs/1954MNRAS.114..437M} {114, 437}

\bibitem[\protect\citeauthoryear{{Peters}, {Klessen}, {Mac Low}  \&
  {Banerjee}}{{Peters} et~al.}{2010}]{2010Peters}
{Peters} T.,  {Klessen} R.~S.,  {Mac Low} M.-M.,   {Banerjee} R.,  2010,
  \mn@doi [\apj] {10.1088/0004-637X/725/1/134}, \href
  {https://ui.adsabs.harvard.edu/\#abs/2010ApJ...725..134P} {725, 134}

\bibitem[\protect\citeauthoryear{{Sana} et~al.,}{{Sana} et~al.}{2012}]{Sana}
{Sana} H.,  et~al., 2012, \mn@doi [Science] {10.1126/science.1223344}, \href
  {http://adsabs.harvard.edu/abs/2012Sci...337..444S} {337, 444}

\bibitem[\protect\citeauthoryear{{Spitzer}}{{Spitzer}}{1978}]{1978.spitzer}
{Spitzer} L.,  1978, {Physical processes in the interstellar medium},
  \mn@doi{10.1002/9783527617722.
}

\bibitem[\protect\citeauthoryear{{Sternberg}, {Hoffmann}  \&
  {Pauldrach}}{{Sternberg} et~al.}{2003}]{2003Sternberg}
{Sternberg} A.,  {Hoffmann} T.~L.,   {Pauldrach} A.~W.~A.,  2003, \mn@doi
  [\apj] {10.1086/379506}, \href
  {http://adsabs.harvard.edu/abs/2003ApJ...599.1333S} {599, 1333}

\bibitem[\protect\citeauthoryear{{Str{\"o}mgren}}{{Str{\"o}mgren}}{1939}]{1939stromgren}
{Str{\"o}mgren} B.,  1939, \mn@doi [\apj] {10.1086/144074}, \href
  {http://adsabs.harvard.edu/abs/1939ApJ....89..526S} {89, 526}

\bibitem[\protect\citeauthoryear{{Tan}, {Beltr{\'a}n}, {Caselli}, {Fontani},
  {Fuente}, {Krumholz}, {McKee}  \& {Stolte}}{{Tan}
  et~al.}{2014}]{2014.Tan.Review}
{Tan} J.~C.,  {Beltr{\'a}n} M.~T.,  {Caselli} P.,  {Fontani} F.,  {Fuente} A.,
  {Krumholz} M.~R.,  {McKee} C.~F.,   {Stolte} A.,  2014, \mn@doi [Protostars
  and Planets VI] {10.2458/azu_uapress_9780816531240-ch007}, \href
  {http://adsabs.harvard.edu/abs/2014prpl.conf..149T} {pp 149--172}

\bibitem[\protect\citeauthoryear{{Toomre}}{{Toomre}}{1964}]{toomre}
{Toomre} A.,  1964, \mn@doi [\apj] {10.1086/147861}, \href
  {http://adsabs.harvard.edu/abs/1964ApJ...139.1217T} {139, 1217}

\bibitem[\protect\citeauthoryear{{Vandenbroucke} \& {Wood}}{{Vandenbroucke} \&
  {Wood}}{2018}]{BV2018}
{Vandenbroucke} B.,  {Wood} K.,  2018, \mn@doi [Astronomy and Computing]
  {10.1016/j.ascom.2018.02.005}, \href
  {http://adsabs.harvard.edu/abs/2018A%26C....23...40V} {23, 40}

\bibitem[\protect\citeauthoryear{{Verner} \& {Ferland}}{{Verner} \&
  {Ferland}}{1996}]{Verner}
{Verner} D.~A.,  {Ferland} G.~J.,  1996, \mn@doi [\apjs] {10.1086/192284},
  \href {http://adsabs.harvard.edu/abs/1996ApJS..103..467V} {103, 467}

\bibitem[\protect\citeauthoryear{{Wolfire} \& {Cassinelli}}{{Wolfire} \&
  {Cassinelli}}{1987}]{wolfire.cassinelli}
{Wolfire} M.~G.,  {Cassinelli} J.~P.,  1987, \mn@doi [\apj] {10.1086/165503},
  \href {http://adsabs.harvard.edu/abs/1987ApJ...319..850W} {319, 850}

\bibitem[\protect\citeauthoryear{{Wood}, {Mathis}  \& {Ercolano}}{{Wood}
  et~al.}{2004}]{KW2004}
{Wood} K.,  {Mathis} J.~S.,   {Ercolano} B.,  2004, \mn@doi [\mnras]
  {10.1111/j.1365-2966.2004.07458.x}, \href
  {http://adsabs.harvard.edu/abs/2004MNRAS.348.1337W} {348, 1337}

\bibitem[\protect\citeauthoryear{{Wood}, {Barnes}, {Ercolano}, {Haffner},
  {Reynolds}  \& {Dale}}{{Wood} et~al.}{2013}]{2013Wood}
{Wood} K.,  {Barnes} J.~E.,  {Ercolano} B.,  {Haffner} L.~M.,  {Reynolds}
  R.~J.,   {Dale} J.,  2013, \mn@doi [\apj] {10.1088/0004-637X/770/2/152},
  \href {http://adsabs.harvard.edu/abs/2013ApJ...770..152W} {770, 152}

\bibitem[\protect\citeauthoryear{{Zinnecker} \& {Yorke}}{{Zinnecker} \&
  {Yorke}}{2007}]{zinnecker}
{Zinnecker} H.,  {Yorke} H.~W.,  2007, \mn@doi [\araa]
  {10.1146/annurev.astro.44.051905.092549}, \href
  {http://adsabs.harvard.edu/abs/2007ARA%26A..45..481Z} {45, 481}

\makeatother
\end{thebibliography}




\bsp	
\label{lastpage}
\end{document}